\documentclass[fleqn]{2020SCGE}
\usepackage{color}
\usepackage{float}
\usepackage{rotfloat}
\usepackage{rotating}
\usepackage{epsfig}
\usepackage{booktabs}
\usepackage{amsmath}
\usepackage{graphicx}
\usepackage{natbib,times}
\usepackage{amsmath}
\usepackage{url}
\usepackage{hyperref}
\usepackage{pdflscape}

\def\kms{km~s$^{-1}$}
\def\HI{H\textsc{i} }
\def\HII{H\textsc{ii} }

\begin{document}
	
	\ensubject{subject}

	
	\ArticleType{Article}
	\SpecialTopic{Special Topic: Peering into the Milky Way by FAST}
	\Year{2022}
	\Month{December}
	\Vol{65}
	\No{12}
	\DOI{10.1007/s11433-022-2040-8}
	\ArtNo{129702}
    \ReceiveDate{July 14, 2022}
    \AcceptDate{November 17, 2022}
	\OnlineDate{November 23, 2022}
	
	\title{Peering into the Milky Way by FAST: \\  I. Exquisite \HI structures in the inner Galactic disk from the piggyback line observations of the FAST GPPS survey}{Peering into the Milky Way by FAST:  I. Exquisite \HI structures in the inner Galactic disk from the piggyback line observations of the FAST GPPS survey}
	
	
	\author[1]{Tao Hong}{hongtao@nao.cas.cn}
	\author[1,2]{JinLin Han}{hjl@nao.cas.cn}
	\author[1]{LiGang Hou}{}
	\author[1,2]{XuYang Gao}{}
	\author[1]{Chen Wang}{}
	\author[1,2]{Tao Wang}{}
	%

\footnotetext[1]{$*$ Corresponding authors (Tao Hong, email: hongtao@nao.cas.cn; JinLin Han, email: hjl@nao.cas.cn)}

	
    \AuthorMark{T. Hong}
	\AuthorCitation{T. Hong, J. L. Han, L. G. Hou, X. Y. Gao, C. Wang and T. Wang}

	\address[{\rm1}]{National Astronomical Observatories, Chinese Academy of
		Sciences, Beijing {\rm 100101}, China}
	\address[{\rm2}]{School of Astronomy, University of Chinese Academy of Sciences, Beijing {\rm 100049}, China}

	\abstract{Neutral hydrogen (H\textsc{i}) is the fundamental component of the interstellar medium. The Galactic Plane Pulsar Snapshot (GPPS) survey is designed for hunting pulsars by using the Five-hundred-meter Aperture Spherical radio Telescope (FAST) from the visible Galactic plane within $|b| \leq 10^{\circ}$. The survey observations are conducted with the L-band 19-beam receiver in the frequency range of 1.0 $-$ 1.5~GHz, and each pointing has an integration time of 5 minutes. The piggyback spectral data simultaneously recorded during the FAST GPPS survey are great resources for studies on the Galactic \HI distribution and ionized gas. We process the piggyback \HI data of the FAST GPPS survey in the region of $33^{\circ} \leq l \leq 55^{\circ}$ and $|b| \leq 2^{\circ}$. The rms of the data cube is found to be approximately 40~mK at a velocity resolution of $0.1$ \kms, placing it the most sensitive observations of the Galactic \HI by far. {The high velocity resolution and high sensitivity of the FAST GPPS \HI data enable us to detect weak exquisite \HI structures in the interstellar medium.} \HI absorption line with great details can be obtained against bright continuum sources. 
	The FAST GPPS survey piggyback \HI data cube will be released and updated on the web: {\it
			\color{blue}http://zmtt.bao.ac.cn/MilkyWayFAST/}.
	}%
	
\keywords{Key Words: interstellar medium, atoms, radio lines, surveys}
	
\PACS{98.38.Gt, 95.30.Ky, 95.80.+p, 95.85.Bh\\}
	
\maketitle
	
	
\begin{multicols}{2}
\section{Introduction}           
\label{sec:intro}
		
\begin{table*}[!t]
	\renewcommand\arraystretch{0.92}
	\newcommand{\tabincell}[2]{\begin{tabular}{@{}#1@{}}#2\end{tabular}}
	\caption[]{Basic parameters of famous Galactic \HI surveys }  \small
	\label{tab:comp}
	\centering
	\resizebox{\textwidth}{!}{
	\begin{tabular}{lcccccc}
	\hline
	\hline
	 Survey & Ref. & Sky  & Spatial resolution & Velocity range & Velocity resolution & Sensitivity \\
	 Name   &  &coverage  &  & (\kms)  & (\kms)  & (mK)   \\
	\hline
	\multicolumn{7}{c}{Interferometer surveys} \\
	\hline
	VGPS & [1] &\tabincell{c}{$18^{\circ} < l < 67^{\circ}$ \\ $|b| < 1.3^{\circ}$ to $2.3^{\circ}$} & $1'$ & ($-113$, +166)  & 0.82  & 2\,000 \\
	GASKAP-\HI & [2] & \tabincell{c}{$167 \leq l \leq 79^{\circ}$, $|b| < 10^{\circ}$; \\ $+$Magellanic Stream and Clouds}&$30''$ & ($-268$, +335) & 0.98 & 1\,100\\
	CGPS & [3] & \tabincell{c}{$74.2^{\circ} < l < 147.3^{\circ}$ \\ $-3.6^{\circ} < b < 5.6^{\circ}$} & $1'$ & ($-165$, +60) & 0.82 & 3\,000\\
	SGPS & [4] & \tabincell{c}{$5^{\circ} < l < 20^{\circ}$, $253^{\circ} < l < 358^{\circ}$ \\ $|b| < 1.5^{\circ}$} & $\sim 3.3'$ &($-227$,  +265) & 0.8  & 1\,600\\
	THOR-\HI & [5] & \tabincell{c}{$14.0^{\circ} < l < 67.4^{\circ}$ \\ $|b| < 1.25^{\circ}$} & $40''$ & ($-139$, +139) & 1.5 & 4\,000\\
	\hline
	\multicolumn{7}{c}{Single dish telescope surveys} \\
	\hline
	LAB & [6] & full sky& $36'$ & ($-400$, +400) & 1.25 & 80\\
	EBHIS & [7] & $\delta \geq -5 ^\circ$ & $10.8'$ & ($-600$, +600) & 1.44  & 90 \\
	GASS & [8] & $\delta \leq 1 ^\circ$ & $16.2'$ & ($-470$, +470) & 1.00  & 57 \\
	HI4PI & [9] & full sky & $16.2'$ & ($-600$, +600) & 1.49  & 43 \\
	GALFA-\HI & [10] & $-1^\circ \leq \delta \leq 38 ^\circ$ & $4'$ & ($-650$, +650) & 0.18  & 352 \\
	{\bf GPPS-\HI} & this work & \tabincell{c}{$\sim30^{\circ} < l < \sim97^{\circ}$, $\sim145^{\circ} < l < \sim215^{\circ}$ \\ $|b| \leq 10^{\circ}$} & {\bf 2.9$'$} & ($-300$, +300) & {\bf 0.10} &  {\bf 40}\\
	\hline
	\end{tabular}}	
	\begin{tablenotes}
	
	\item[1] \textbf{Reference:} 
	[1]: \citet{Stil2006}; [2]: \citet{Dickey2013}; [3]: \citet{Taylor2003}; [4]: \citet{McClure-Griffiths2005};  [5]: \citet{Wang2020}; [6]: \citet{Kalberla2005}; [7]: \citet{Winkel2016}; [8]: \citet{Kalberla2010}; [9]: \citet{HI4PI2016}; [10]: \citet{Peek2018}.
    \end{tablenotes}
\end{table*}

Neutral hydrogen (H\textsc{i}) is one of the key components of matters in spiral galaxies. It is widely distributed in the Milky Way. 
The \HI gas is the raw material for star formation and plays an important role in galaxy evolution \citep[][and reference therein]{Kulkarni1987, Kalberla2009}. Observing the emission line of the Galactic \HI gas can reveal the structure and dynamics of the Galaxy over a much extended region in the Galactic disk even in the very far side of the Galaxy. Furthermore, observing the Galactic \HI gas is necessary for understanding the life cycles in the Galactic interstellar medium \citep[ISM; ][]{Klessen2016}.

\begin{figure*}[!t]
	\centering
	\includegraphics[width=0.81\textwidth]{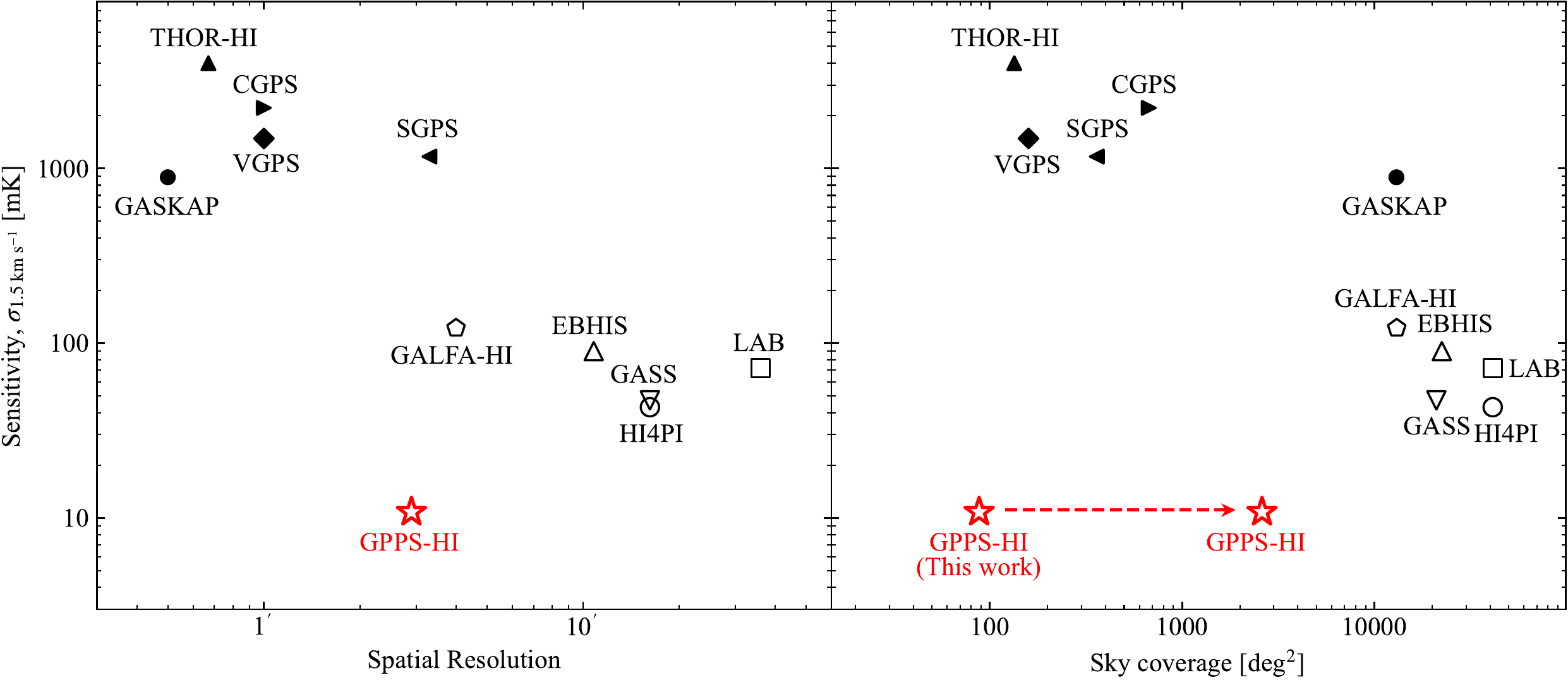}
	\caption{Comparison of the normalized flux sensitivity $\sigma_{\rm 1.5~km~s^{-1}}$ at the same spectral resolution of 1.5 \kms\, for the Galactic \HI surveys versus the survey spatial resolution ({\it left panel}, the best in the bottom left corner) and the sky coverage ({\it right panel}, the best in the bottom right corner). The interferometric surveys are labeled with the filled symbols and the single-dish surveys by open symbols. Reference:  GASKAP-\HI: \citet{Dickey2013}; THOR-\HI: \citet{Wang2020}; CGPS: \citet{Taylor2003};  VGPS: \citet{Stil2006}; SGPS: \citet{McClure-Griffiths2005}; GALFA-\HI: \citet{Peek2018}; EBHIS: \citet{Winkel2016}; GASS: \citet{Kalberla2010}; HI4PI: \citet{HI4PI2016};	LAB: \citet{Kalberla2005}; GPPS-\HI: this work.}
    \label{fig:comp_survey}
\end{figure*}

Both single dish radio telescopes and interferometers have been used to observe the Galactic \HI gas (see Table~\ref{tab:comp}). Interferometric \HI observations \citep[e.g. ][]{McClure-Griffiths2005, Taylor2003} can have a high spatial resolution but with a relatively low sensitivity. The VLA Galactic Plane Survey \citep[VGPS; ][]{Stil2006} is one of the best representatives. It has a high angular resolution of $\sim 1'$ with a sensitivity of 2~K per 0.824 \kms\,channel. The ongoing Galactic ASKAP Survey \citep[GASKAP;][]{Dickey2013, Pingel2022, Dickey2022} improves the spatial resolution to $30''$ with a rms noise of 1.1~K per 0.98 \kms. Single-dish Galactic \HI surveys have a much higher sensitivity and cover a much wider region in the sky. One of the most widely-used single-dish Galactic \HI surveys is the Leiden/Argentine/Bonn Survey \cite[LAB; ][]{Kalberla2005}. This survey was conducted by several 25-m class telescopes, and provides a sensitivity of 80~mK per 1.25 \kms\,channel with a coarser spatial resolution of $36'$. With multi-beam receivers, the survey efficiency is largely improved, and several modern \HI surveys have been accomplished in recent years. Using the 100-m Effelsberg telescope with a seven-beam 1.4-GHz (L-band) receiver, the Effelsberg-Bonn \HI Survey \citep[EBHIS,][]{Winkel2016} observed the entire northern hemisphere with a sensitivity of 90~mK per 1.44 \kms\,channel. GASS \citep{McClure-Griffiths2009, Kalberla2010, Kalberla2015} mapped the southern sky with a sensitivity of 57~mK per 1.00 \kms\,channel by using the L-band 13-beam receiving system mounted on the Parkes 64-m radio telescope. By combining the data of EBHIS and GASS, the HI4PI \citep{HI4PI2016} covers the full sky with a sensitivity of approximately $43$~mK and a velocity resolution of 1.49 \kms, while the spatial resolution is only $16.2'$. The Galactic Arecibo L-Band Feed Array \HI \citep[GALFA-H\textsc{i}; ][]{Peek2011, Peek2018} covers the Arecibo sky with a resolution of $4'$ and a r.m.s of $352$~mK per 0.18 \kms\,channel. As summarized in Fig.~\ref{fig:comp_survey}, the interferometric surveys (filled black symbols) have a high spatial resolution but a low flux sensitivity, while the single-dish surveys (open symbols) have a wider sky coverage and a higher sensitivity, but with a lower spatial resolution.

\HI is a good tracer of Galactic filaments and spiral arms \citep{Burton1971, Levine2006, Koo2017}. By adopting the unsharp masking method, \citet{Kalberla2016} studied the local Galactic \HI distribution and revealed small-scale \HI structures with combined data from GASS and EBHIS.  
\citet{Soler2020} searched for \HI filaments in the Galaxy with the Hessian matrix method and found a long \HI filament named ``Maggie" with a length of 1.2~kpc using the data of the H\textsc{i}/OH/Recombination line survey of the inner Milky Way \citep[THOR-H\textsc{i};][]{Wang2020}, and \citet{Soler2022} further applied the method to the HI4PI survey \citep{HI4PI2016} and found a large number of filamentary structures even in the very outer Galaxy. By using the \HI data observed by the Five-hundred-meter Aperture Spherical radio Telescope (FAST) together with data from the HI4PI, \citet{Li2021} detected an \HI filament located behind the Outer Scutum–Centaurus (OSC) arm, which might be the farthest \HI filament in the Galaxy.
		
By observing the spectral line signal against strong continuum sources, the \HI 21cm line is shown as absorption features. The strong background sources could be extra-galactic radio sources or strong continuum sources within the Milky Way, such as \HII regions or supernova remnants (SNRs).  Combining \HI absorption observations with observation results from molecular lines, one can estimate kinematic distances of SNRs \citep[e.g.][]{Tian2008, Ranasinghe2018}. The \HI self-absorption (HISA) traces the cold neutral medium (CNM) which absorbs \HI emission from warmer background gas \citep{Gibson2000, Gibson2005}. \HI narrow-line self-absorption \citep[HINSA,][]{Li2003} is a special case of HISA, which is caused by the cold \HI that exists around the molecular clouds. \citet{Tang2020} and \citet{Liu2022} observed \HI toward selected Planck Galactic cold clumps by using FAST and detected HINSA features in 58\% and 92\% of their samples, respectively.
		
FAST is by far the largest single-dish radio telescope in the world \citep{Nan2011}. It provides the highest sensitivity at L-band for pulsar and spectroscopy observations by using the 19-beam receiver \citep{Jiang2020}. The Galactic Plane Pulsar Snapshot (GPPS) survey\footnote{see 
  { \color{blue}http://zmtt.bao.ac.cn/GPPS/ }for survey details, including the sky coverage, beam arrangement and current progress.} \citep{Han2021} is designed for hunting pulsars in the FAST accessible Galactic plane within $|b| \leq 10^\circ$. During the pulsar survey observations, high resolution spectral data are simultaneously recorded in a piggyback mode using another set of backend, covering the band from 1.0 to 1.5~GHz with 1024~K channels. With an integration time of 5 minutes for each pointing, the sensitivity of \HI observations can reach approximately $40$~mK at a velocity resolution of 0.1 \kms, which is by far the most sensitive Galactic \HI observation (see Table~\ref{tab:comp} and Figure~\ref{fig:comp_survey}).  
		
This series of papers are dedicated to investigate the interstellar medium by FAST. 
This is the first paper for \HI gas. The second paper is on the ionized gas by \citet{Hou2022}, also coming from the piggyback spectral data from the GPPS survey.  
A study of the Galactic magnetic fields in a deep interstellar space is presented by \citet{Xu2022} in the third paper, using mainly the pulsar Faraday rotation measures observed in the GPPS survey. The identification of two large supernova remnants from the FAST-scanned continuum radio maps is presented by \citet{Gao2022} in the fourth paper.
		
In this paper, we process the piggyback \HI data from the FAST GPPS survey, and present the HI map for a sky area of 88 square degrees in the inner Galactic disk. The GPPS survey strategy and the spectral data are briefly introduced in Section~\ref{sec:obs}. We describe the data processing in Section~\ref{sec:data}, and the \HI results for the sky area of $ 33^{\circ} \leq l \leq 55^{\circ}$, $|b|\leq 2^{\circ}$ and some highlights are presented in Section~\ref{sec:release}. We summarize the results in Section~\ref{sec:summary}.

\section{The GPPS survey and piggyback spectra}
\label{sec:obs}
		
To hunt pulsars, the FAST GPPS survey \citep{Han2021} observes the Galactic plane using the L-band 19-beam receiver. The observations are carried out with a full gain of the FAST in originally down to $26.4^{\circ}$ from the zenith but now extended to $28.5^{\circ}$, within which the FAST has an almost full illumination aperture of 300~m. The snapshot observational mode is developed by \citet{Han2021} for doing four pointings of the 19 beams (M01 to M19) for 5 minutes each, and they successively slew the beams, one after another, with a separation of $3'$ in between. The beam size at 1420 MHz is $2.82'$ to $2.95'$, as the beams at the edges of the 19-beam receiver have a slightly larger beam size and stronger side-lobes \citep[see table 2 of][]{Jiang2020}. The beam-switch takes less than 20~s each, and four 
pointings of 19 beams take 20 minutes, so that in 21 minutes, a hexagonal sky area of 0.1575 square degree can be fully covered by $4\times19 = 76$ beams \citep[see Fig. 4 in][]{Han2021}. There is almost no overlap for these 76 beams within the extent of the full width half maximum (FWHM), so that the sky coverage is not Nyquist-sampled in terms of beam size. In total 18\,438 snapshot covers are expected to be carried out for the entire GPPS survey, and about 1\,800 covers have been observed so far. The detailed observational method and survey strategy of the GPPS survey can be found in \citet{Han2021} and the progress can be found on the GPPS web-page\footnotemark[1].

\begin{table}[H]
	\caption[]{Parameters for the FAST GPPS survey and piggyback \HI spectra}
	\renewcommand\arraystretch{0.92}
	\newcommand{\tabincell}[2]{\begin{tabular}{@{}#1@{}}#2\end{tabular}}
	\small
	\label{tab:basic_para}
	\centering
	\begin{tabular}{ll}
		\hline
		\hline
		Parameter & Value \\
		\hline
		FAST aperture diameter & 300 m \\
		Beam size & $ \sim 2.9'$ \\
		System temperature & $ \sim $~23~K \\
		Gain &  $ \sim 16~\rm{K~Jy^{-1}}$ \\
		Integration time per pointing & 297 s \\
		Galactic longitude range &  \tabincell{c}{$\sim 30^{\circ} < l < \sim97^{\circ}$, \\$\sim 145^{\circ} < l < \sim 215^{\circ}$}\\
		Galactic latitude range & $|b| \leq 10^{\circ}$ \\
		\hline
		Frequency range & 1000 MHz $-$ 1500 MHz \\
		Channel number & {1024 K} \\
		Frequency resolution & 476.8 Hz \\
		Velocity resolution for \HI & $0.1$ \kms \\
		Sensitivity ($\sigma_{\rm 0.1~km~s^{-1}}$) & {40~mK} \\
		\hline
	\end{tabular}
\end{table}
		
The GPPS survey records the spectral data together with the pulsar data simultaneously using a separate spectroscopy backend during observations. The major parameters of the GPPS survey and 
the \HI data are summarized in Table~\ref{tab:basic_para}. 
		
The FAST L-band 19-beam receiver can receive radio signals in the frequency range of 1000~MHz to 1500~MHz, and they are channelized into {1024~K channels} in the digital backend. The four polarization products XX, YY, XY* and YX* are accumulated for a sampling time of 1~s, and recorded simultaneously during the GPPS survey observations for each beam. To calibrate the receiving system, a reference signal with a temperature amplitude of about 1K was injected to the system every other second. A few 2-min cal-on and cal-off data-takings are conducted in the beginning or the end or in the middle of an observation session for a given day. The data recorded with calibration on-off signal are very useful to calibrate and convert the data unit to the antenna temperature in Kelvin. 
		
\section{Processing piggyback \HI data of the FAST GPPS survey}
\label{sec:data}
		
A few steps are needed to process the piggyback \HI data of FAST GPPS survey, as described in the following.
		
\subsection{Data preparations}
		
One snapshot observation consists of four 5-minute pointings. During the observations, the spectroscopy digital backend records the data consecutively without stops but generates a series of fits files, and each contains 128 one-second samplings. Hence for each beam in the snapshot observation of four pointings, ten fits files are written in the FAST data repository for one snapshot observation, including about 1\,252 data samplings. 
		
The first step of the data processing is to cut and then combine the ten spectrum files on the basis of the observation time for four pointings. For each beam, the data recorded during the source switching time are discarded, and finally four spectrum files corresponding to the four pointings in the snapshot are prepared, each associated with one pointing position. For 19 beams with four pointings, one gets 76 files for a snapshot observation.
		
In principle, every pointing for each beam should have on-source time of exactly 300~s. In practice, data shortage for 1~s or 2~s occasionally occurs at the very beginning or the end due to either radio frequency interference (RFI) or other telescope operation restrictions, and the real integration time is always about 297 seconds. 
		
\subsection{Calibration on brightness temperature }
\label{sec:cali}
		
As mentioned above, in the observation session of a day, two or three or even four data segments are recorded for two minutes each with calibration signals on-off. The digital difference between the averages for the on and off phases, corresponding to the reference level of 1~K temperature in $T_{a}$, gives the scale of the snapshot data that day. We apply the scale to the data to convert the received power into antenna temperature $T_{a}$. As described by \citet{Jiang2020}, the electronic gain of the FAST system fluctuates less than 1\% over 3.5 hours, the band-pass varies about $4\%$ in 30 minutes. Our observation sessions typically last less than 3 hours, the differences between the scales we obtained at the beginning and the end of an observation session are generally less than 5\%. The averaged value of the conversion scales over one observation session is applied to minimize the effects of system fluctuations.
		
Brightness temperature $T_{b}$, however, is more often used in studying the Galactic \HI emission. To convert the antenna temperature $T_{a}$ to the brightness temperature $T_{b}$, the main beam efficiency is needed. Through the measurements toward the calibrator 3C~138, \citet{Gao2022} obtained the FAST gain and the beam-widths for the FAST L-band 19-beam receiver. The main beam efficiency $T_{a}/T_{b}$ is then determined, as shown in their figure~2. We apply their results at 1420~MHz to our \HI data. Note that the values given by \citet{Gao2022} are derived from the eight observations toward 3C~138 during December 2020 and March and April 2021, and that the GPPS data were taken in a much longer period from 2019 to a recent date. Therefore applying the $T_{a}/T_{b}$ converting factor of \citet{Gao2022} to all the GPPS spectral line data is not ideal but a necessary step. Fortunately our results can be further verified by the EBHIS data (see Sect.~\ref{sec:cal_ebhis}).
		
\subsection{RFI mitigation}
		
The GPPS survey piggyback \HI data are very occasionally affected by narrow- and wide-band radio frequency interference. The narrow-band RFI occasionally affects tens of channels (several \kms\,in the velocity space) of a spectrum, which we can mask manually and replace them by using a linear interpolation of their neighbours. The wide-band RFI extends several hundreds of \kms. It has to be fitted and removed together with the standing wave and baseline by the routine described in Sect.~\ref{sec:baseline} below.

\subsection{Standing waves and baseline}
\label{sec:baseline}
		
The radio wave reflections between the main dish and the receiver cabin of a telescope cause periodic fluctuations in the spectra, which is known as the ``standing waves". The standing wave frequency of FAST is about 1 MHz, corresponding to a velocity width of about $200$~\kms\, for the \HI line. The standing waves are one of the main nuisances of extra-galactic \HI observations of FAST, but do not severely affect the observations of the Galactic \HI signals, because the local \HI signal is obviously stronger and wider. However, to obtain the accurate flux density and reveal the weak \HI structures, standing waves must be removed. 
		
A traditional method for removing the standing waves is to fit the line-free area of the spectrum by using a sine function, and then subtracting the fitted model from the data. However, we find that the standing waves of the GPPS \HI spectrum vary with both time and frequency. A single sine function often fails in fitting the standing waves in the spectrum range of wider than $1\,000$~\kms.
		
\begin{figure}[H]
	\centering
	\includegraphics[width=0.4\textwidth]{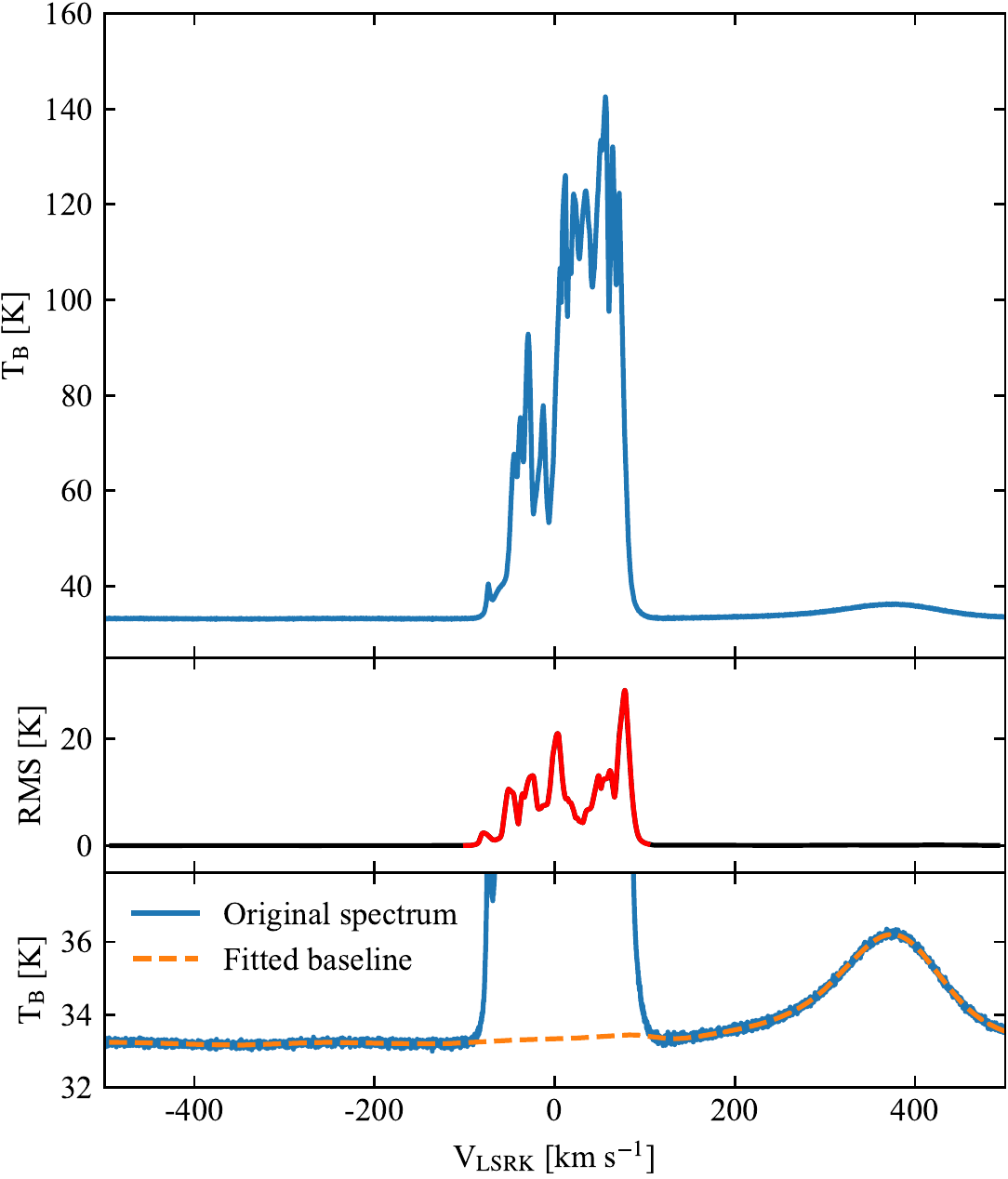}
	\caption{Example for the \HI line finding and the baseline subtraction. {\it Top panel} is the original \HI spectrum; {\it Middle panel} shows the rms variation curve used to identify the \HI line region, calculated from every 160 adjacent channels in the original \HI spectrum shown in the top panel; {\it Bottom panel} is the original spectrum (blue solid line) together with the fitted baseline and standing waves (orange dashed line)  by the ArPLS \citep{Baek2015, Zeng2021}.}
	\label{fig:baseline}
\end{figure}

A new routine which automatically fits the standing waves and the baseline together is developed in this work. The first step is to locate the \HI line region in the frequency channels by calculating the \textit{rms} of every 160 adjacent channels in the spectrum. The \HI line region is identified with a threshold over $3\sigma$ level on this \textit{rms} variation curve. The \HI line identified through the \textit{rms} data, instead of the original spectrum line, can avoid the mis-identification of the wide band RFI efficiently. An example is shown in Fig.~\ref{fig:baseline}. The \textit{rms} curve shown in the {\it middle panel} is calculated from the original spectrum in the {\it top panel}. The red line part marks the detected line region. This routine identifies the \HI line accurately and ignores the wide band RFI in the velocity range of $200$~\kms\,to 500~\kms. The standing waves can then be fitted together with the baseline. 

\begin{figure*}[!t]
	\centering
	\includegraphics[width=0.83\textwidth]{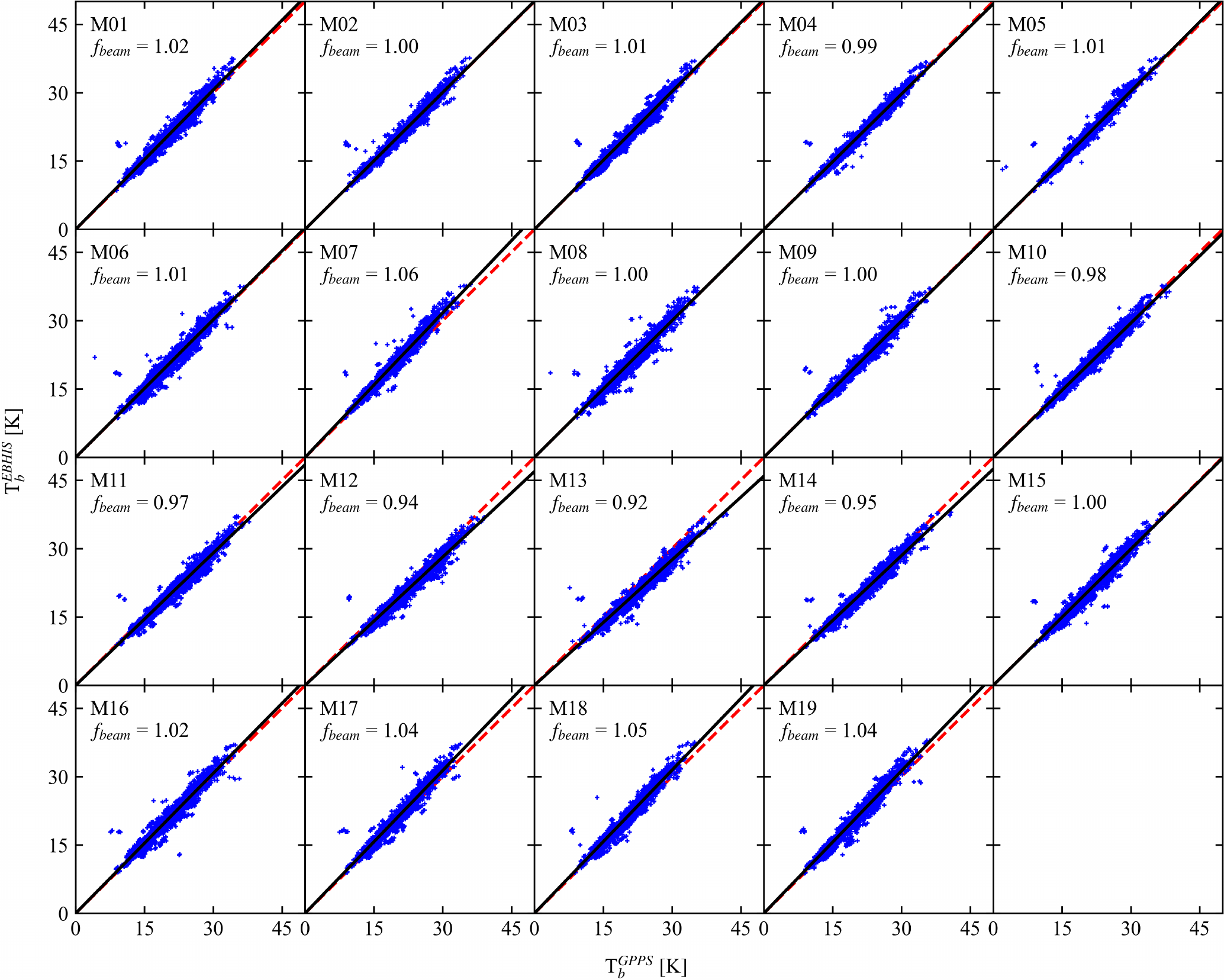}
	\caption{Comparison of the GPPS mean brightness temperatures with those data of the EBHIS for each of 19 beams (beam No. M01 to M19) respectively yield the gain correction factor $f_{\rm beam}$ for each beam. The solid lines present the best fitting relations to the data, the dashed lines stand for the equality.}
	\label{fig:tt_plot_1}
\end{figure*}

The second step is to apply the asymmetrically re-weighted penalized least squares smoothing method \citep[ArPLS, ][]{Baek2015, Zeng2021} to the line-free region. The ArPLS method provides fast and accurate estimates to the baseline of the \HI spectra. The fitting result stands for not only the standing waves but also other irrelevant features such as the continuum emission even in the range of the \HI line, which is then subtracted from the \textit{original} spectrum. Therefore the RFI, standing waves and the continuum emission have been all well removed. This approach works successfully on almost all spectra, only occasionally fails in a few cases because of strong RFI, for which we re-fit the baseline manually.
		
\begin{figure}[H]
	\centering
	\includegraphics[width=0.4\textwidth]{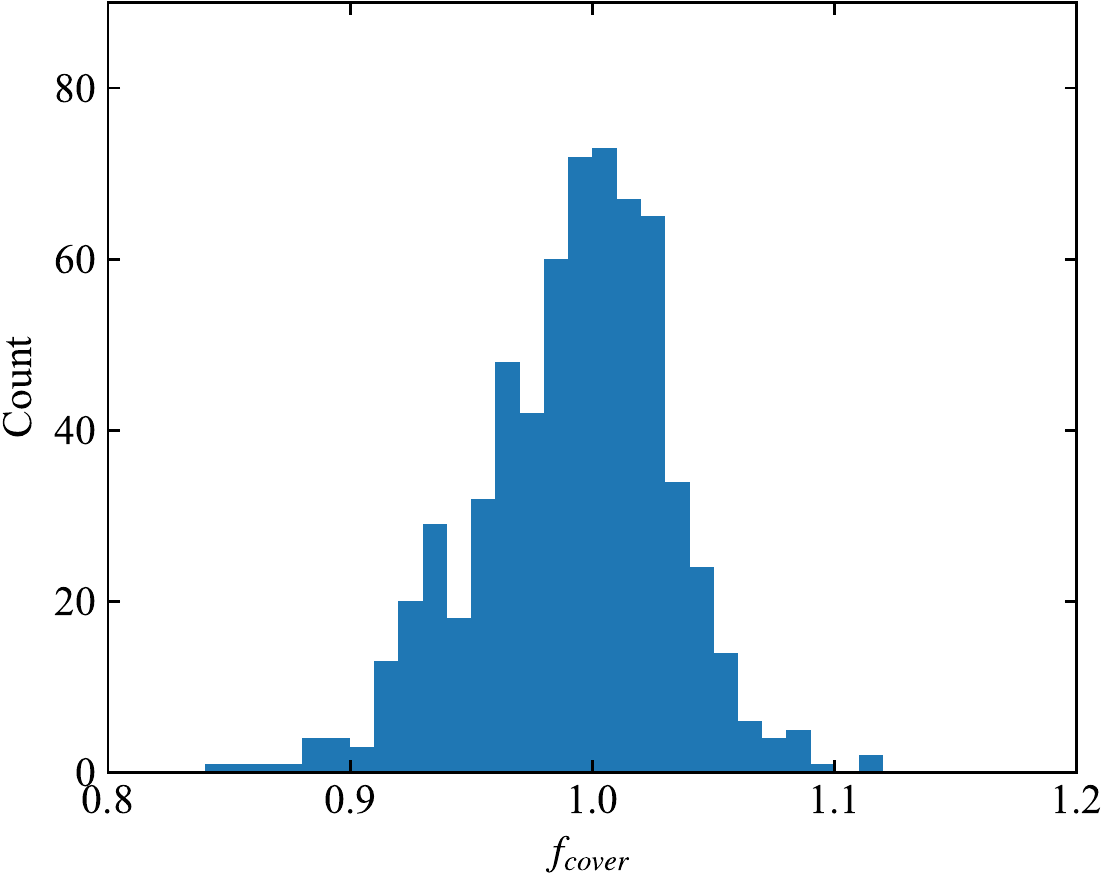}
	\caption{The number distribution of the cover gain calibration factor  $f_{cover}$.}
	\label{fig:factor}
\end{figure}

\begin{sidewaysfigure*}[tp]
  \vspace{9.5cm} 
	\centering 
	\includegraphics[width=0.85\textheight]{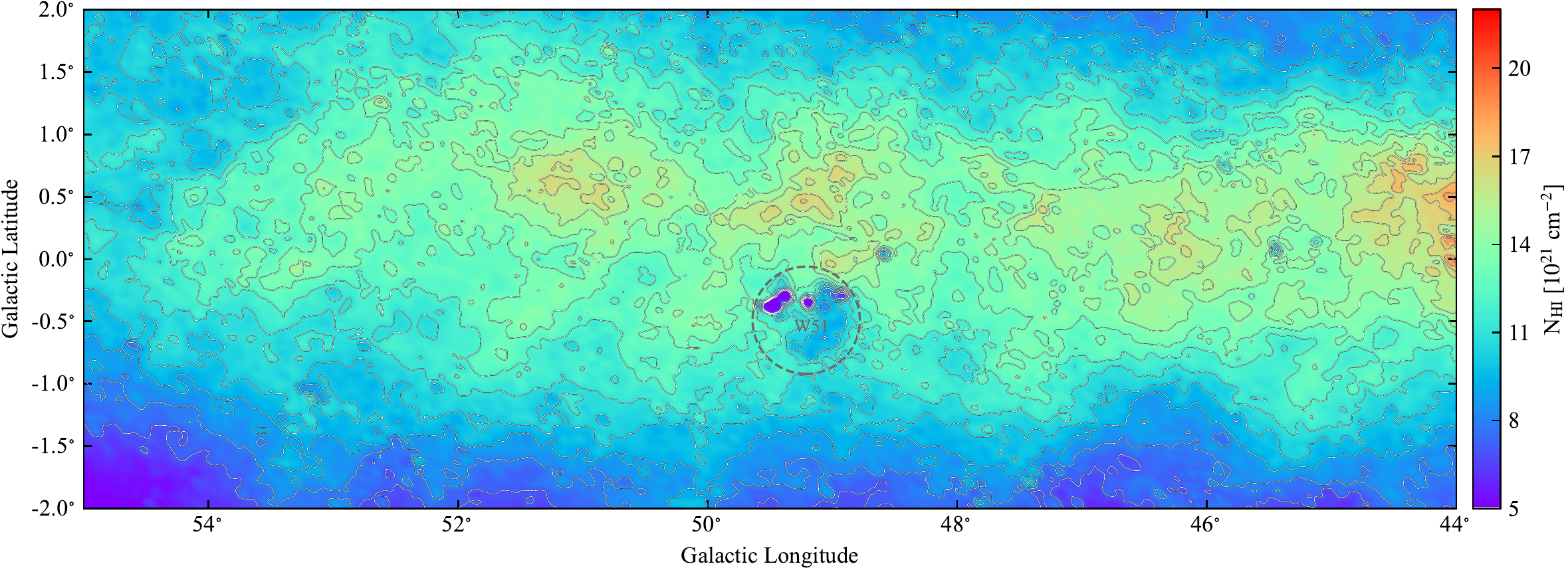} \\
	\vspace{0.8cm}
	\includegraphics[width=0.85\textheight]{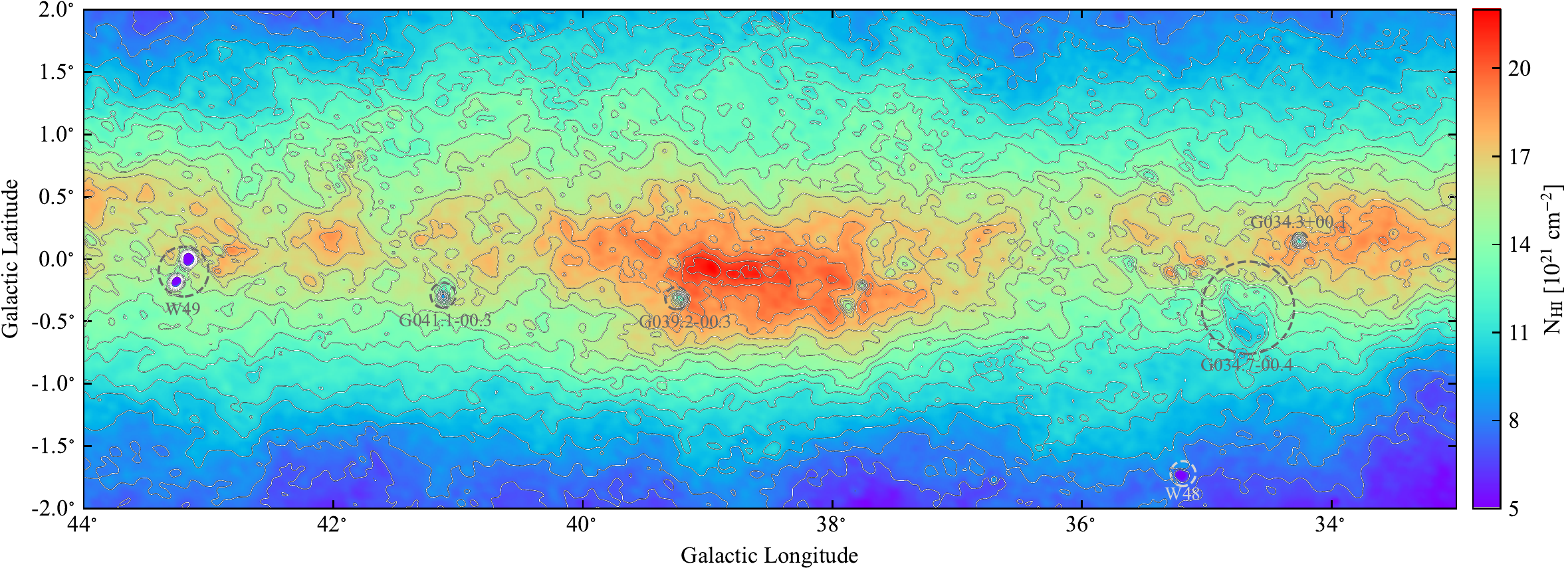}
	\caption{The \HI column density map integrated over the velocity range of $-150$~\kms\,to $150$~\kms. The overlaid contour levels are 7 to 21 $\times 10^{21}$ cm$^{-2}$ with a step of 1 $\times 10^{21}$ cm$^{-2}$. Low \HI column density areas correspond to star formation complexes W48, W49 and W51, supernova remnants G034.7$-$00.4, G039.2$-$00.3 and G041.1$-$00.3, and \HII region G034.3+00.1 as marked by dashed circles.}
	\label{fig:final}
\end{sidewaysfigure*}

\begin{figure*}[tp]
	\centering
	\includegraphics[width=0.95\textwidth]{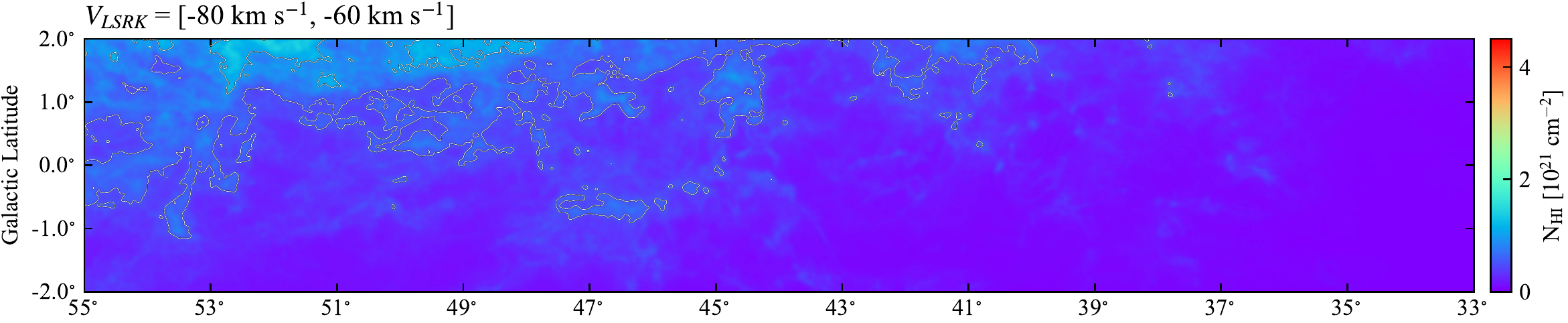}
	\includegraphics[width=0.95\textwidth]{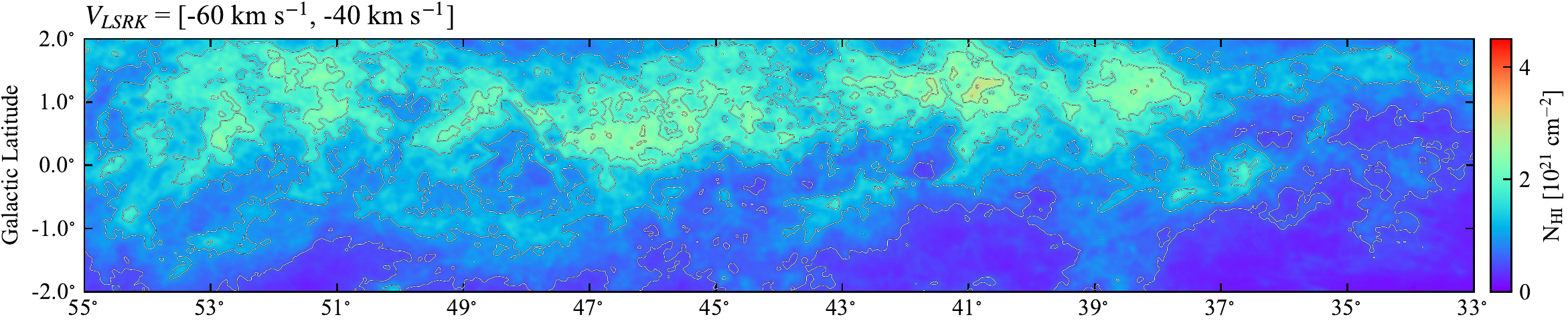}\\
	\includegraphics[width=0.95\textwidth]{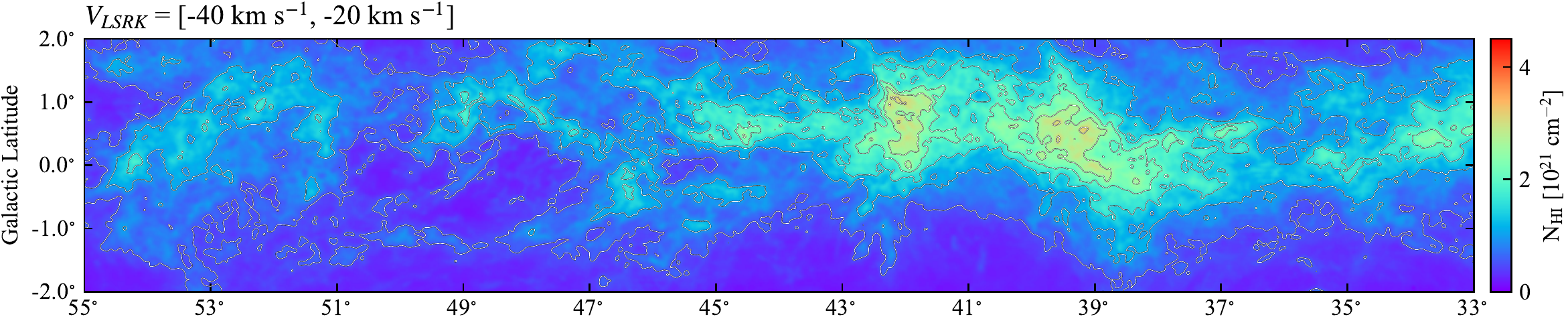}\\
	\includegraphics[width=0.95\textwidth]{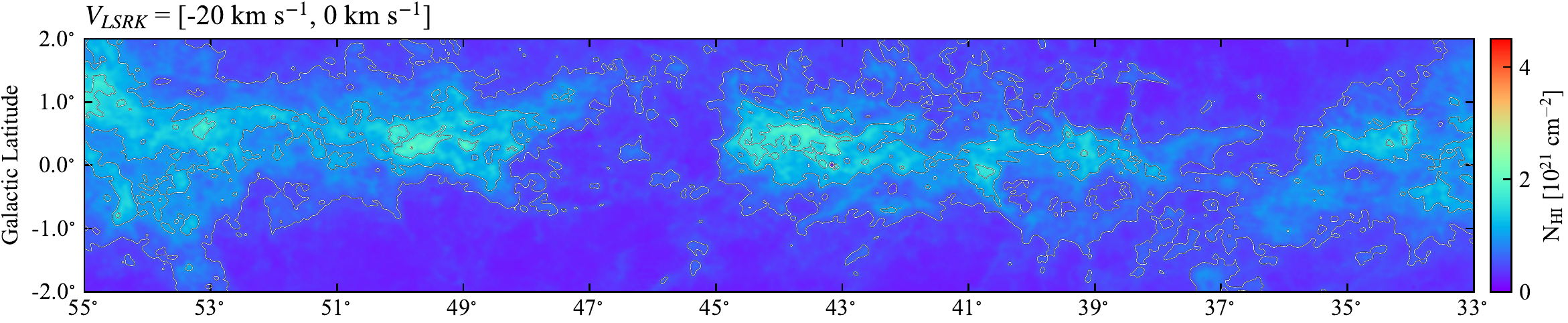}\\
	\includegraphics[width=0.95\textwidth]{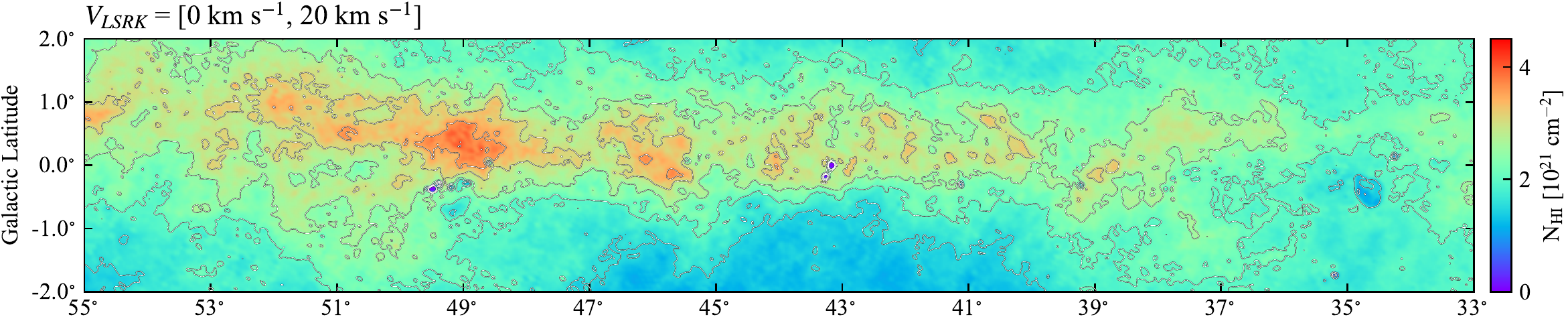}\\
	\includegraphics[width=0.95\textwidth]{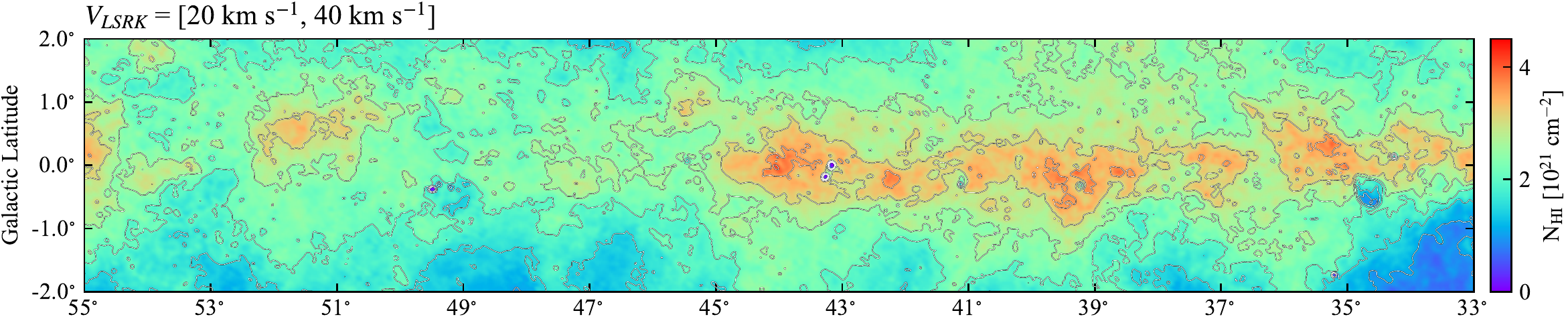}\\
	\caption{Channel maps of the \HI column density in the velocity range of $-80$~\kms\,to $120$~\kms\,with a channel width of $20$~\kms. The overlaid contour levels are 0.5 to 4 $\times 10^{21}$ cm$^{-2}$ with a step of 0.5 $\times 10^{21}$ cm$^{-2}$. --- {\it to be continued } -- }
	\label{fig:chan_map}
\end{figure*}

\addtocounter{figure}{-1}
\begin{figure*}[!t]
	\centering
	\includegraphics[width=0.95\textwidth]{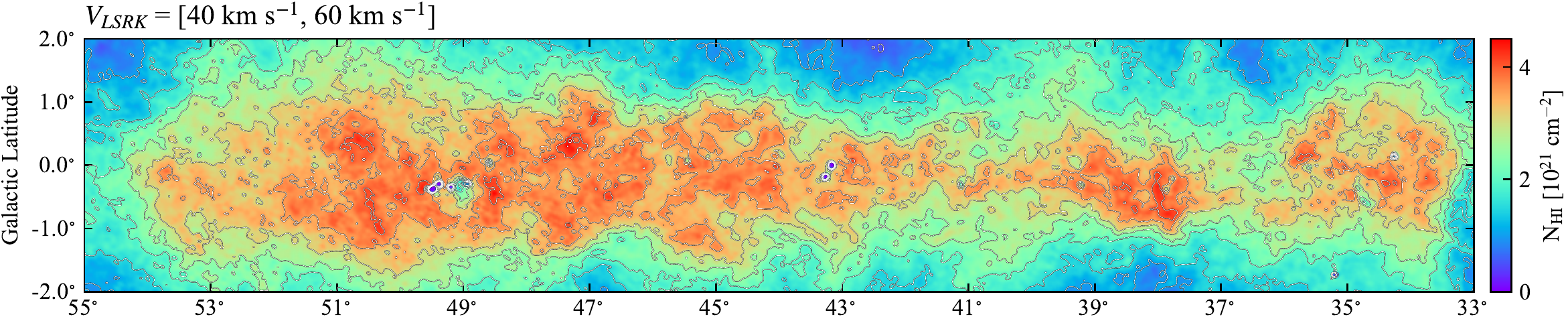}\\
	\includegraphics[width=0.95\textwidth]{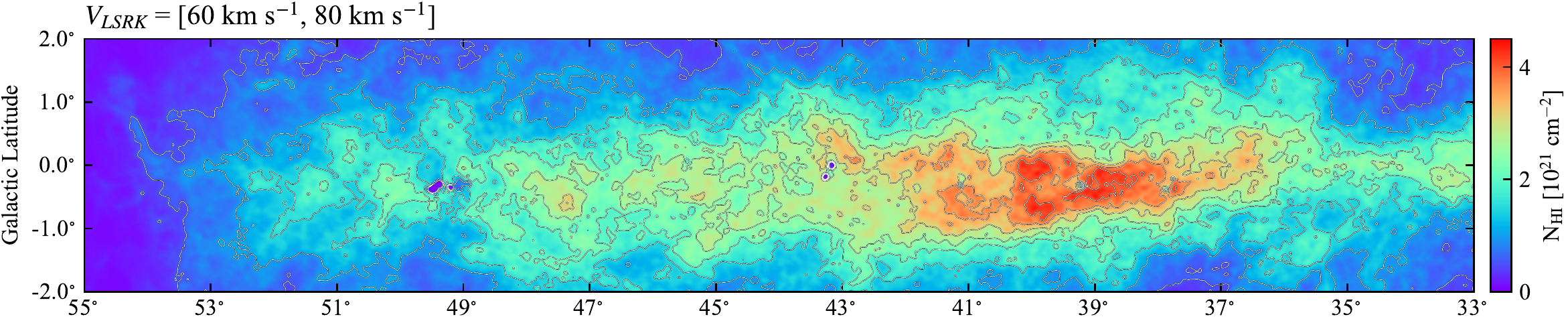}\\
	\includegraphics[width=0.95\textwidth]{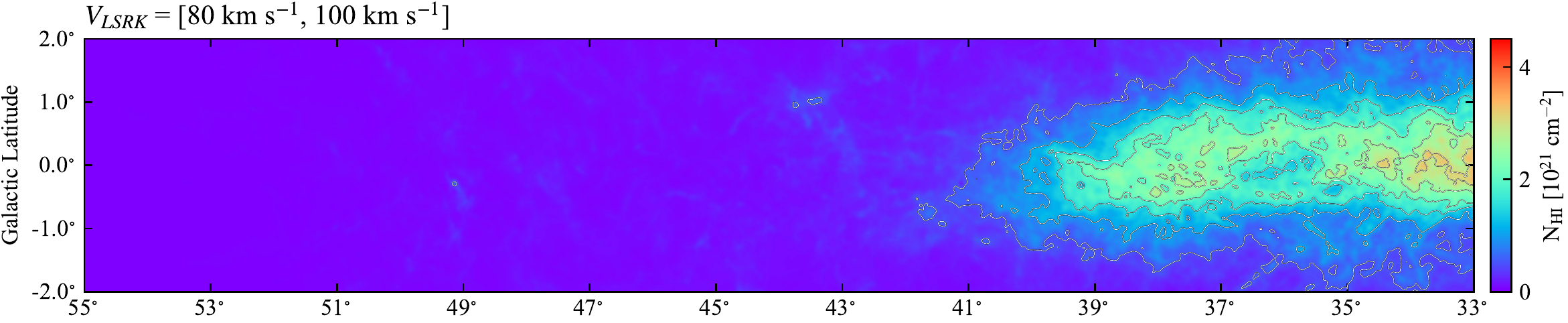}\\
	\includegraphics[width=0.95\textwidth]{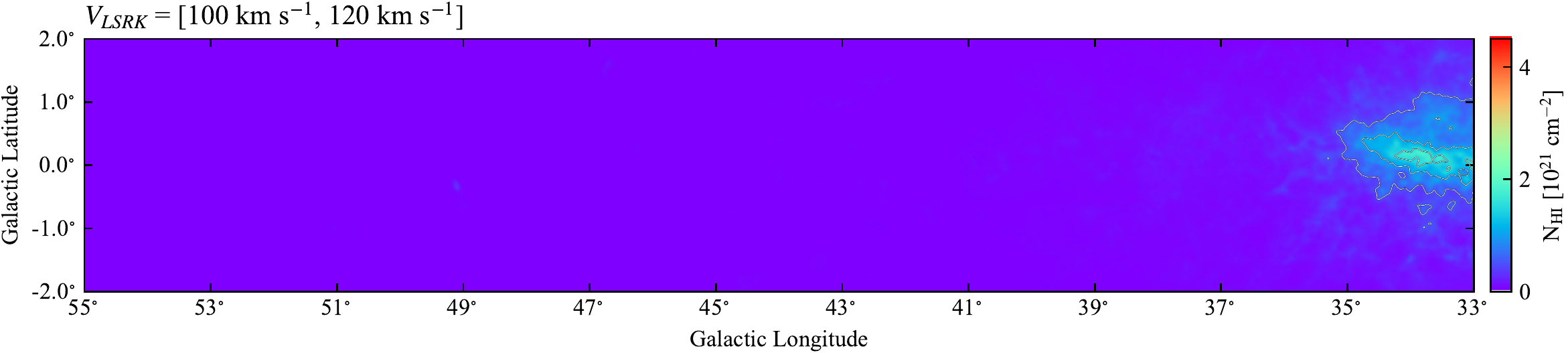}\\
	\caption{--- end.}
	\label{fig:chan_map_2}
\end{figure*}

\begin{figure*}[t]
	\centering
	\includegraphics[width=0.9\textwidth]{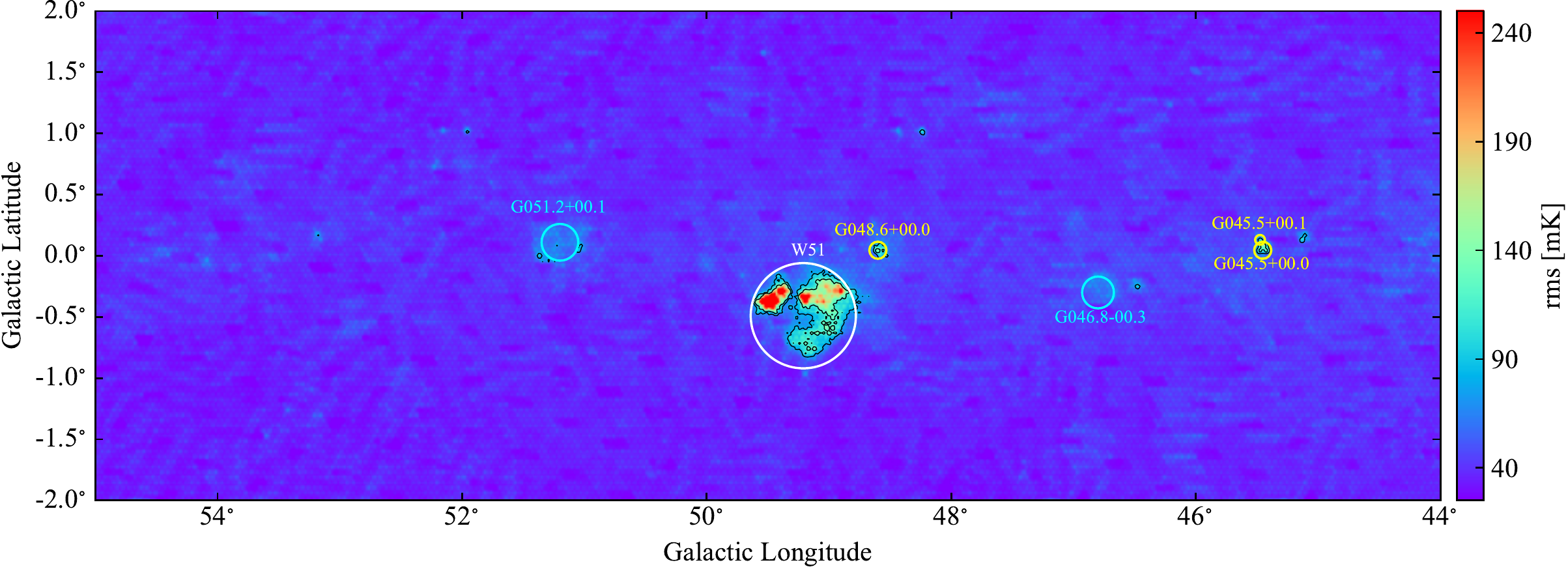}\\
	\includegraphics[width=0.9\textwidth]{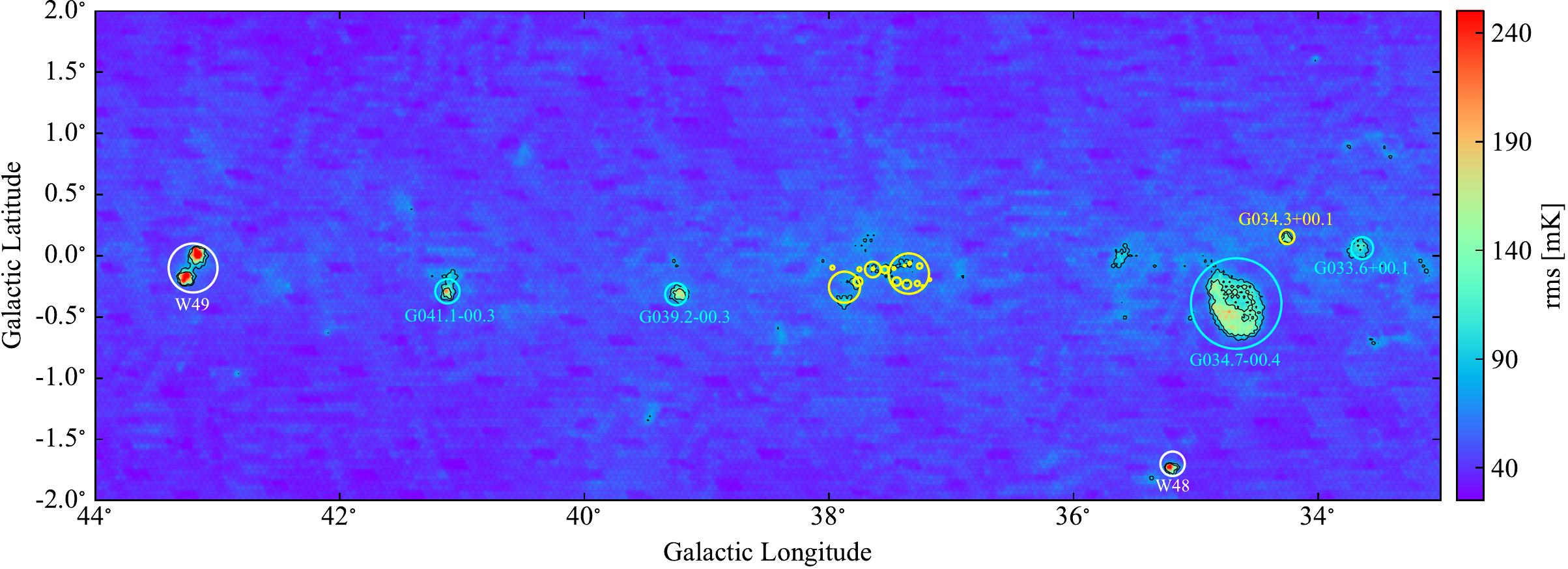}
	\caption{The rms map of the GPPS Galactic \HI data cube with a spatial resolution of $2.9'$ and a velocity resolution of $0.1$~\kms\,in a grid of $1' \times 1'$. The strong continuum emission sources cause relatively high rms, as indicated by circles: star formation complexes W 48, W 49 and W 51 by white circles; supernova remnants G033.6+00.1, G034.7$-$00.4, G039.2$-$00.3, G041.1$-$00.3, G046.8$-$00.3 and supernova remnant candidate G051.2+00.1 by cyan circles; \HII regions G034.3+00.1, G45.5+0.00, G45.5+00.1 and G048.6+00.0 by yellow circles, together the \HII regions in the region of $l = 37.0^\circ$ to $ 38.0^\circ$, $b = -0.4^\circ $ to $ 0.0^\circ$ without labels. The overlaid contour levels are 80~mK and 120~mK, corresponding to the double and triple of the rms value of 40~mK across the area (see Figure~\ref{fig:stats_rms}).}
	\label{fig:rms}
\end{figure*}

\subsection{The stray radiation}

The correction for the stray radiation of different beams has not yet been carried out for the current data set for this initial data release. Fortunately all the GPPS observations presented here are carried out by the main focus receiver with a zenith angle less than $26.4^{\circ}$, so that there is no spillover of the feed illumination beyond the edge of the 500-m primary reflector.

\subsection{Calibrate the GPPS gain for the \HI line to EBHIS}
\label{sec:cal_ebhis}
		
As described above in Sect.~\ref{sec:cali}, the GPPS survey has already accumulated data for a long period of time from 2019 to 2022. However, the conversion factor between $T_{a}$ to $T_{b}$ is derived from the observations toward 3C~138 in December 2020, and March and April in 2021. Therefore the it does not work perfectly for all the survey data obtained in such a long time. The long-term gain variations of the 19-beam receiver have to be calibrated. We compare the GPPS \HI data to the EBHIS data. The EBHIS survey was observed and well calibrated by a fully steerable single-dish telescope with a 7-beam receiver, hence provides a similar antenna status in all directions over the sky, though the EBHIS data has a coarser resolution. 

The GPPS \HI data are calibrated based on the EBHIS results in two steps. The first step is to scale the different gains of 19 beams. We first smooth the GPPS \HI data cube to the same beam size of $10.8'$ as the EBHIS, and then spilt all the data into 19 sub-data sets each corresponding to one FAST beam (i.e. M01 to M19). Comparison between the mean brightness temperature obtained by the GPPS survey and those from the EBHIS in the velocity range of $-150$~\kms\,to 150 \kms\, for these sub-data sets gives a scaling factor from 0.92 to 1.06 for each beam, as shown by the best fitting slopes in Figure~\ref{fig:tt_plot_1}. These values are taken as the correction factors $f_{beam}$ for each beam. After this step, the slightly different gains of the 19 beams near the 1420~MHz are well calibrated.  
		
After gain corrections for different beams, some brightness variations emerge between different snapshot covers. Similarly, we calibrate the gain variations of a given cover as a whole based on the comparison of the smoothed GPPS data with the EBHIS data. The calibration factors $f_{cover}$ are in the range of 0.84 to 1.11, and the distribution of $f_{cover}$ is shown in Fig.~\ref{fig:factor}.

The calibration factors of $f_{beam}$ and 
$f_{cover}$ are applied together to the GPPS data 
to diminish the long time variation of the FAST gain, so that 
$	T_{b}^{\rm corrected} = T_{b}^{\rm obs} \times f_{\rm beam} \times f_{\rm cover}. $
The so-obtained GPPS brightness temperature is then very smooth and agrees with EBHIS data with a mean relative difference of only 1\% (see Sect.~\ref{sec:consist}).

\section{\HI survey results}
\label{sec:release}
		
We process the piggyback spectral data of the GPPS survey \citep{Han2021} and obtain the first part of the Galactic \HI spectra data in the region of $ 33^{\circ} \leq l \leq 55^{\circ}$ and $|b| \leq 2^{\circ}$ in this work. The observations were conducted between March 2019 and October 2022 using the FAST L-band 19-beam receiving system. The results are also available on the web-page\footnote{\color{blue}http://zmtt.bao.ac.cn/MilkyWayFAST/.} for a data cube ($l, b, v$) of the Galactic plane. 
All spectra are Doppler-shifted to the kinematic local standard of rest (LSRK), the velocity range covered by the data cube is $-300$~\kms\,$\leq V_{\rm  LSRK} \leq 300$~\kms\,with a raw velocity resolution of $0.1$~\kms\,and a spatial resolution of $2.9'$. In Fig.~\ref{fig:final}, we show the image of the total \HI column density integrated over the velocity range of $-150$~\kms\,to  $150$~\kms, the range most of the Galactic \HI structures have, by
using the optically thin scaling factor of $1.823 \times 10^{18}~{\rm cm}^{-2} / ({\rm K~km~s}^{-1})$ \citep{Kulkarni1987}. The map is gridded with $1' \times 1'$ pixels. Clearly seen in the map, the 
areas with low \HI column densities correspond to relatively strong continuum emission sources, which often show absorption features on the \HI spectra after the strong continuum emission even in the absorption line range is subtracted in the step of baseline fitting.  
More detailed channel maps in the velocity range of $-80$~\kms\,to $120$~\kms\,are presented in Fig.~\ref{fig:chan_map} with a channel width of $20$~\kms. The \HI structures trace the large-scale mass distribution in the Milky Way, and the \HI distribution in the velocity space roughly relates to the distances of the structures. 
In the \HI images in the velocity range of $-80 $ \kms\,$< V_{\rm  LSRK} < -20$ \kms, the bright \HI emission structures are obviously offset to high Galactic latitudes, which indicates the warp of the outer Galactic disk \citep{Kalberla2009}.

\subsection{Sensitivity assessment}
		
\begin{figure}[H]
	\centering
	\includegraphics[width=0.38\textwidth]{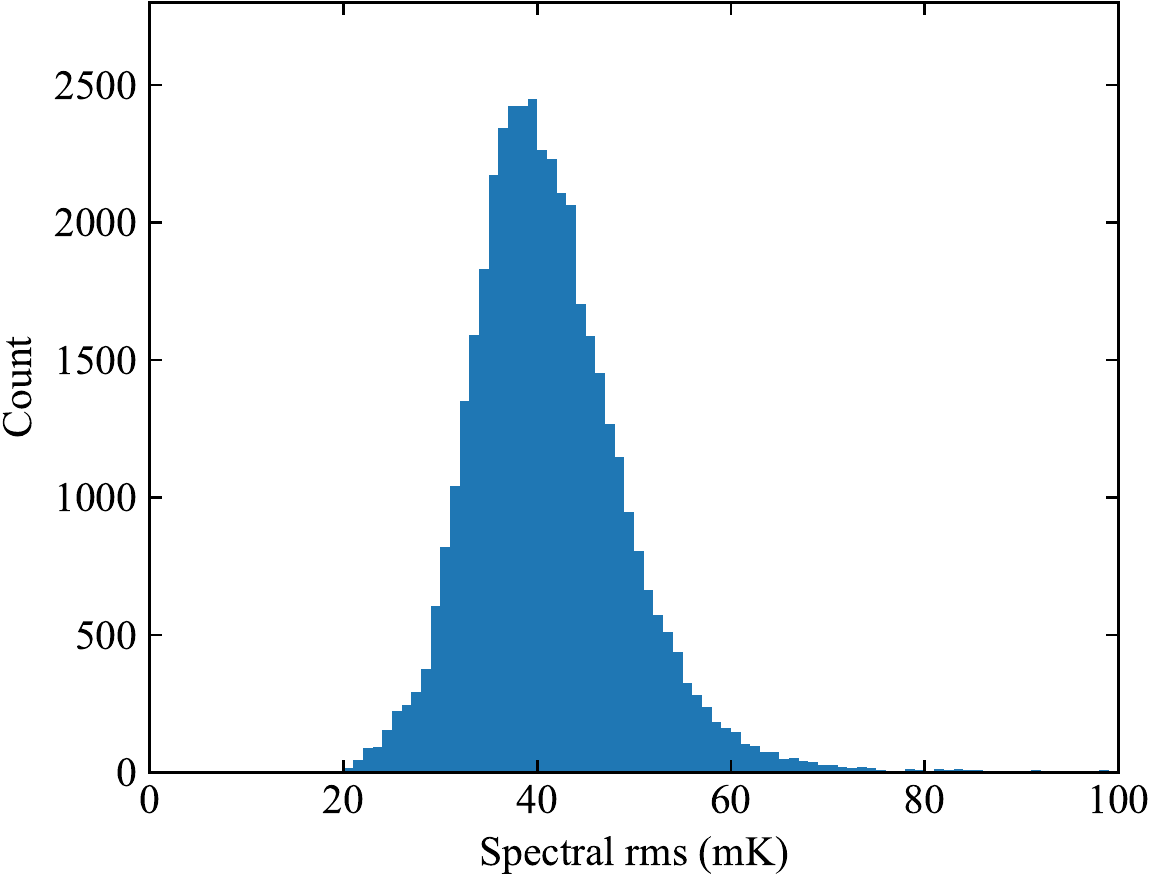}
	\caption{Number distribution of spectral rms values with a velocity resolution of $0.1$~\kms. {The tail end, which is induced by extremely bright continuum sources, contributes a very small fraction of the distribution and is excluded for legibility.} The mean is 41.8~mK, and the median is 40.2~mK. After excluding the tail in the higher end induced by strong continuum sources, the distribution can be fitted with a Gaussian with a peak of $39.9 \pm 6.8$~mK.}
	\label{fig:stats_rms}
\end{figure}

The rms values of the spectra are calculated as the noise of 1\,000 \HI line-free channels at a velocity resolution of $0.1$~\kms\,{(corresponding to the velocity range of $-300$ to $-250$~\kms\,and also the range of +250 to +300~\kms)}, and the rms map is shown in Fig.~\ref{fig:rms}. The distribution of rms values is shown in Fig.~\ref{fig:stats_rms}. The median rms is $40.2$~mK. After the tail of the large value end caused by the strong radio sources is {excluded}, the rms distribution can be fitted with a Gaussian peak at $39.9 \pm 6.8$~mK. Therefore the value of 40~mK is adopted as the sensitivity of the piggyback \HI observations in this section of the FAST GPPS survey.

As shown in Fig.~\ref{fig:rms}, the rms is generally uniform over the surveyed area, except for a few regions of strong continuum sources of star forming regions and supernova remnants. Besides the areas marked in Fig.~\ref{fig:final}, high rms spots are noticed in Fig.~\ref{fig:rms} for the supernova remnants: G033.6+00.1, and G046.8$-$00.3, and \HII regions: G45.5+0.00, G45.5+00.1 and G048.6+00.0. The high rms structure centered at $l = 51.2^\circ$, $b =0.1^\circ$ is a supernova remnant candidate reported by \citet{Anderson2017}. We also notice a high rms area at $l = 37.0^\circ$ to $38.0^\circ$, $b = -0.4^\circ$ to $0.0^\circ$, not only one or two \HII regions but about 20 \HII regions are located in this region \citep{Anderson2014}.
		
\subsection{Comparison with EBHIS and GALFA-\HI}
\label{sec:consist}
		
\begin{figure}[H] 
	\centering
	\includegraphics[width=0.38\textwidth]{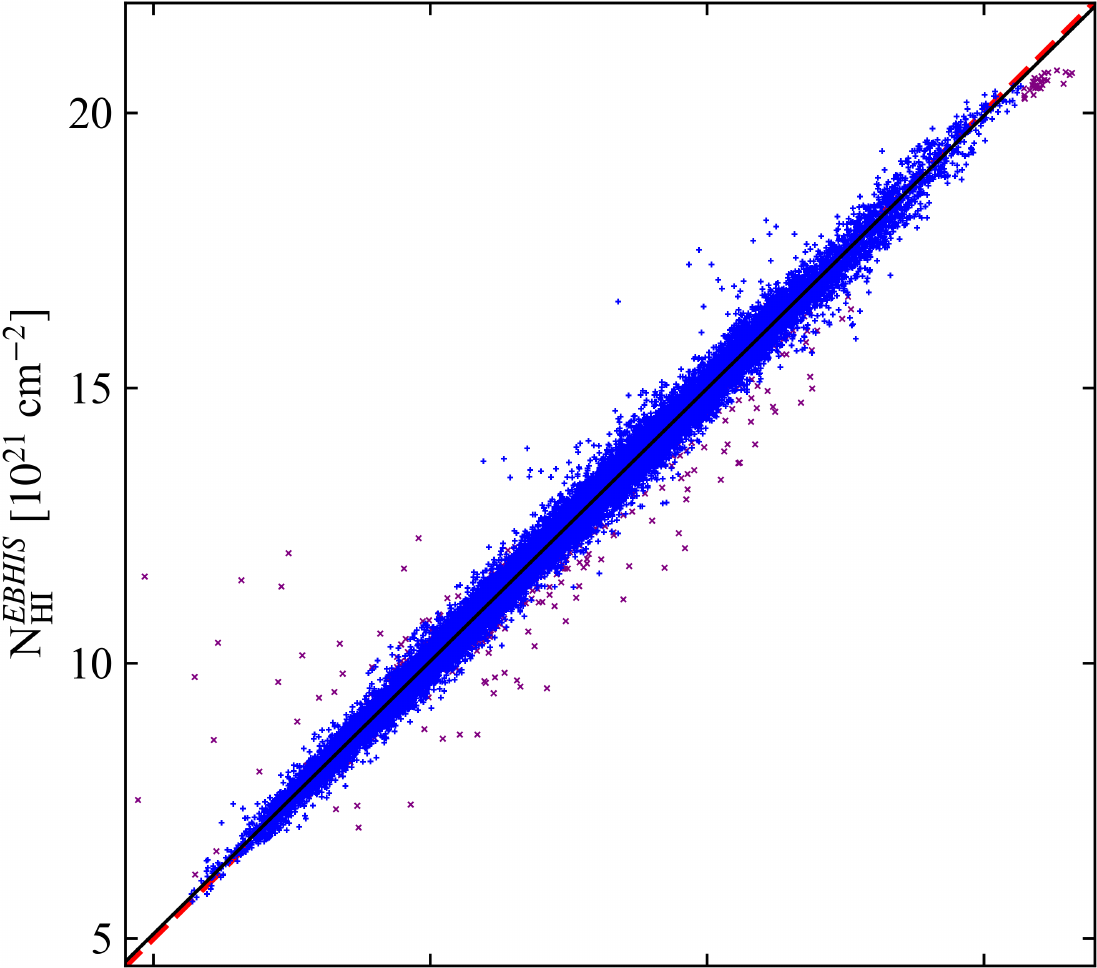} \\
	\includegraphics[width=0.38\textwidth]{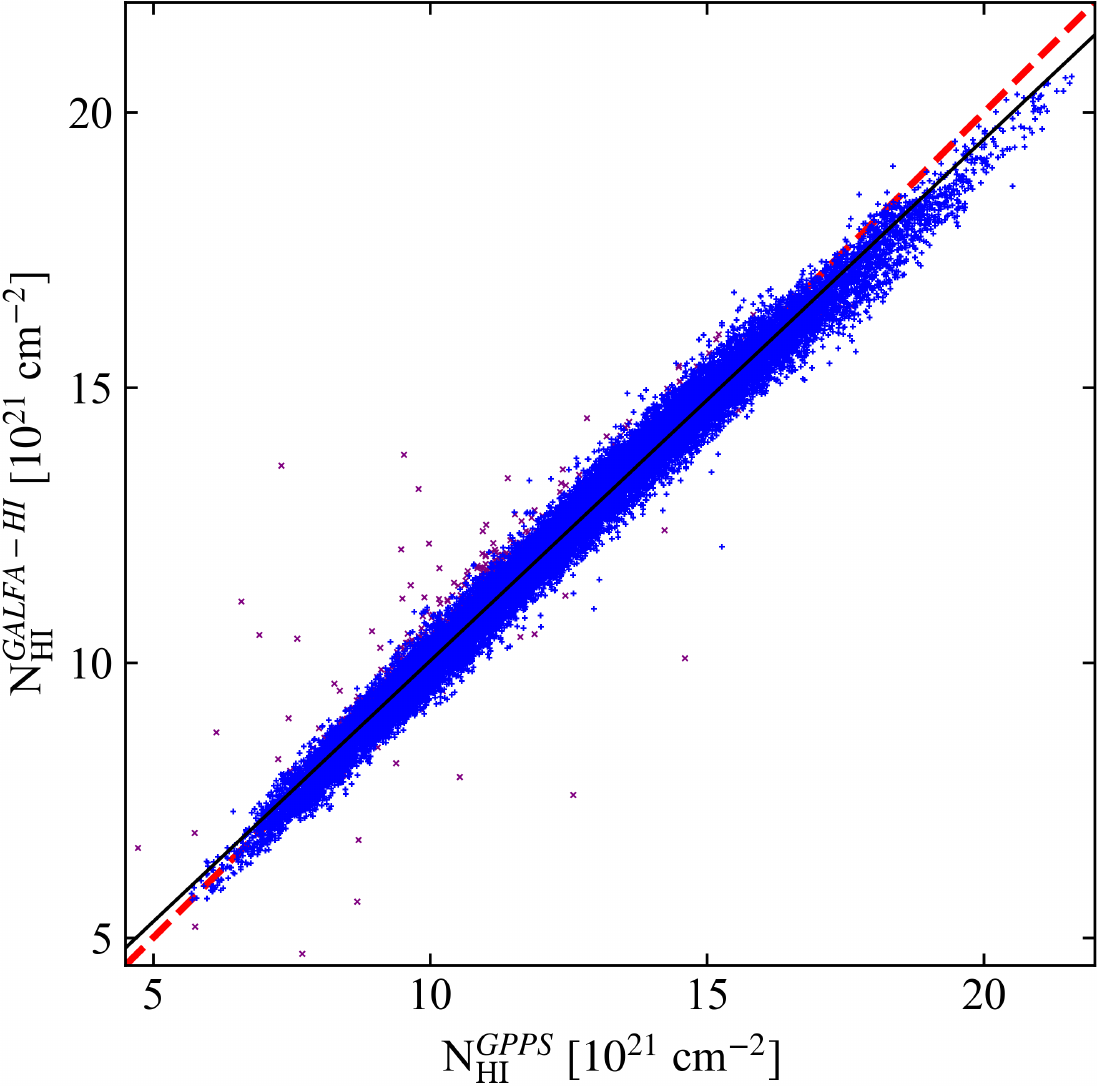}
	\caption{Comparison of the total \HI column densities from the GPPS survey with the EBHIS data ({\it upper panel}) and  GALFA-\HI  ({\it lower panel})  shows a great consistence, with some outliers (shown as little purple `x') due to strong continuum sources. The red dashed line stands for equality, and the black solid line is a linear fitting to the data after excluding outliers.}
	\label{fig:comp}
\end{figure}

We check the consistency of the GPPS \HI column density integrated in the velocity range of $-150$~\kms\,to 150~\kms\,with the EBHIS survey data. The GPPS data cube is smoothed to the same spatial resolution as that of the EBHIS data. The comparison between the two surveys in the upper panel of Fig.~\ref{fig:comp} shows a great consistence, with only a very small number of outliers at the areas of strong radio continuum sources including star forming regions W49 and W51. Once these outliers are excluded, 
the two data sets can be linked with a linear fit as: 
\begin{equation}
	N_{\rm{H\textsc{i}}}^{GPPS} = 1.01 \cdot
	N_{\rm{H\textsc{i}}}^{EBHIS} - 0.11 \times
	10^{21}~\rm{cm}^{-2}.
\end{equation}
Considering the mean value of the \HI column densities $\langle  N_{\rm{H\textsc{i}}}^{GPPS} \rangle = 11.7 \times 10^{21}~\rm{cm}^{-2}$, the zero point offset of $0.11 \times 10^{21}~\rm{cm}^{-2}$ corresponds to a relative offset of only about 1\%. 

The same comparison is made with the GALFA-\HI\,data. We show it in the lower panel of Fig.~\ref{fig:comp} and find
\begin{equation}
	N_{\rm{H\textsc{i}}}^{GPPS} = 1.05 \cdot
	N_{\rm{H\textsc{i}}}^{GALFA-\HI} - 0.58 \times
	10^{21}~\rm{cm}^{-2},
\end{equation}
which indicates that the two data sets agree with each other within 5\%.

\begin{figure*}
	\includegraphics[width=0.28\textwidth]{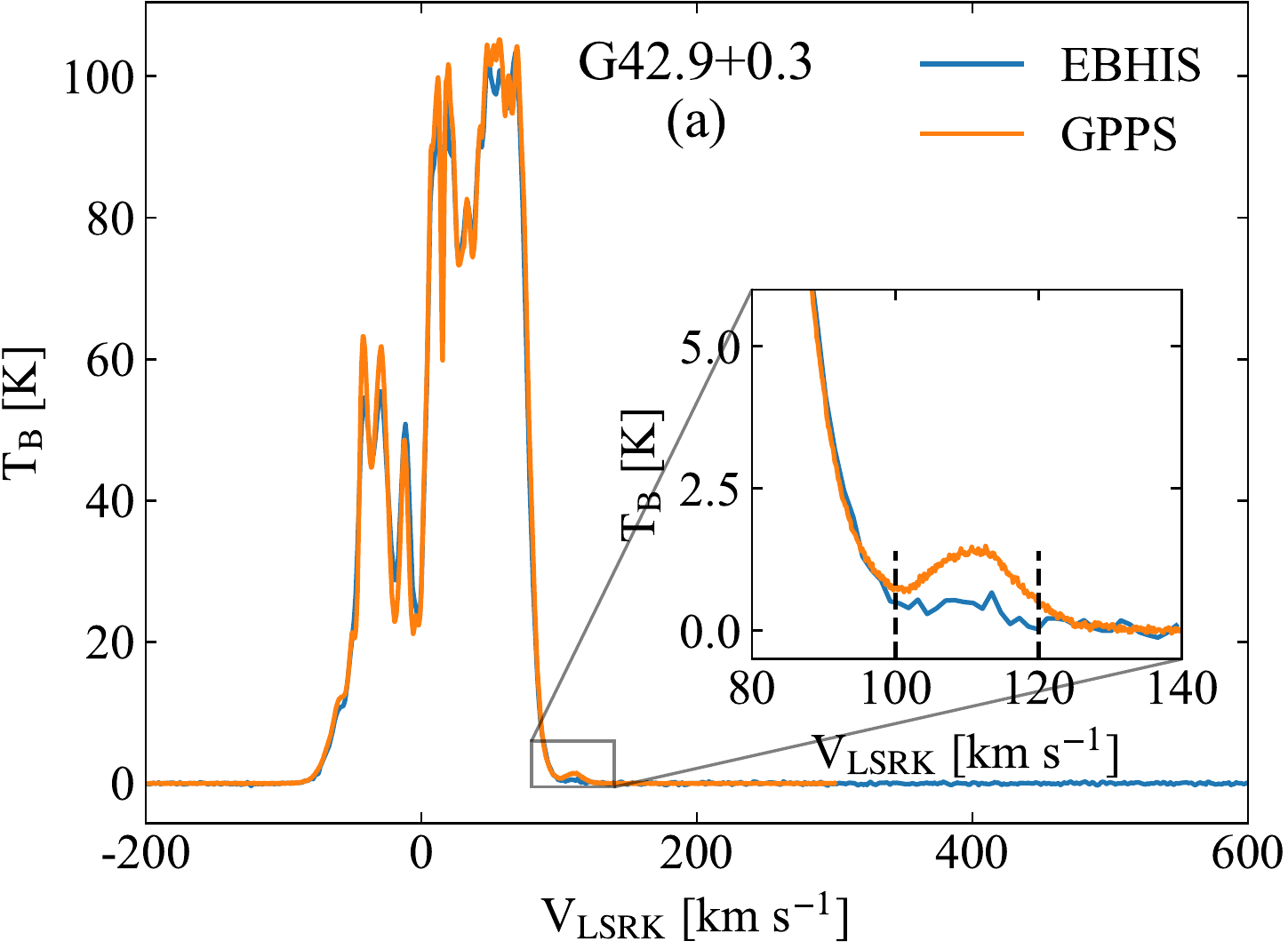}
	\hspace{0.02\textwidth}
	\includegraphics[width=0.28\textwidth]{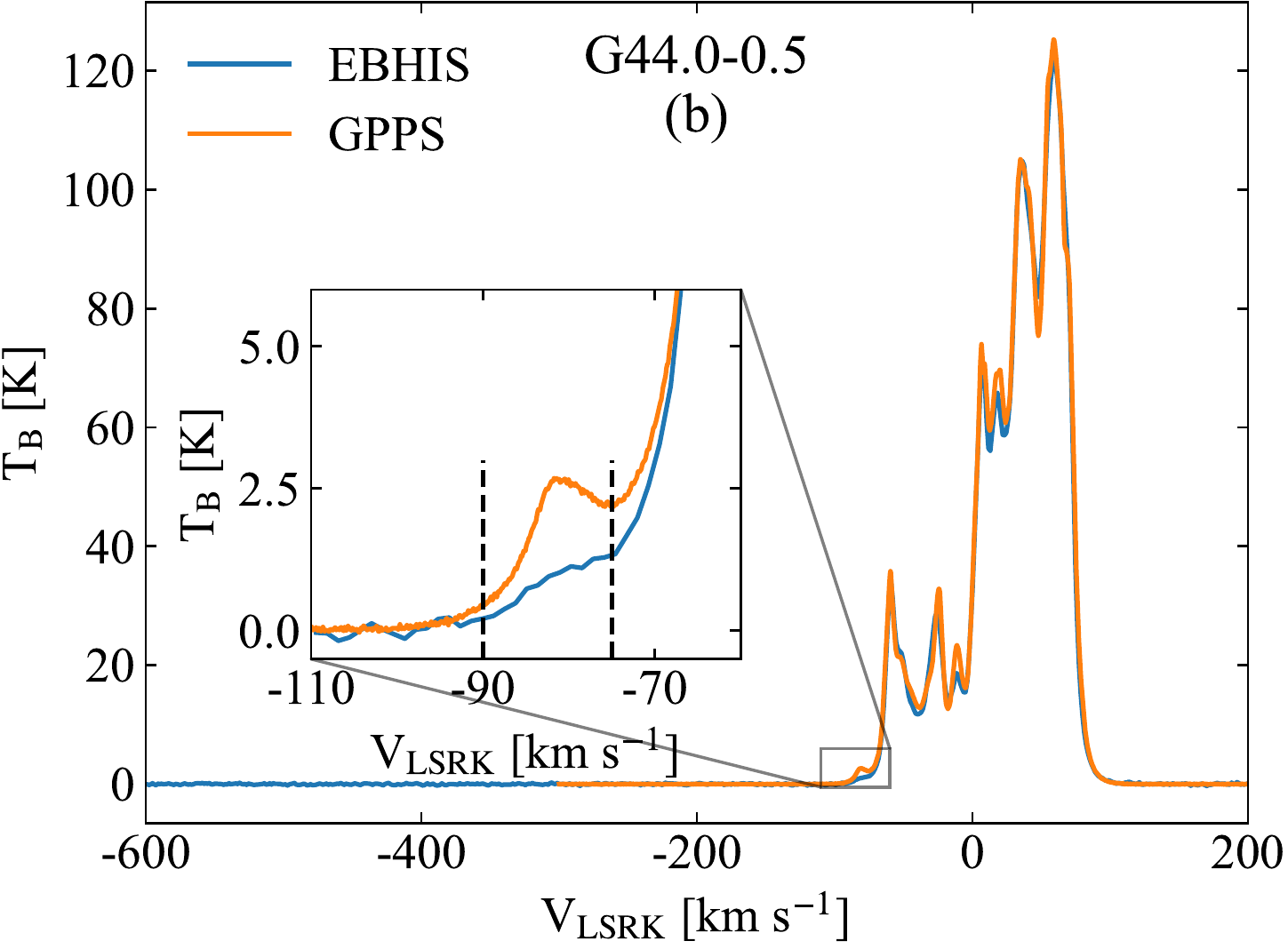}
	\hspace{0.02\textwidth}
	\includegraphics[width=0.28\textwidth]{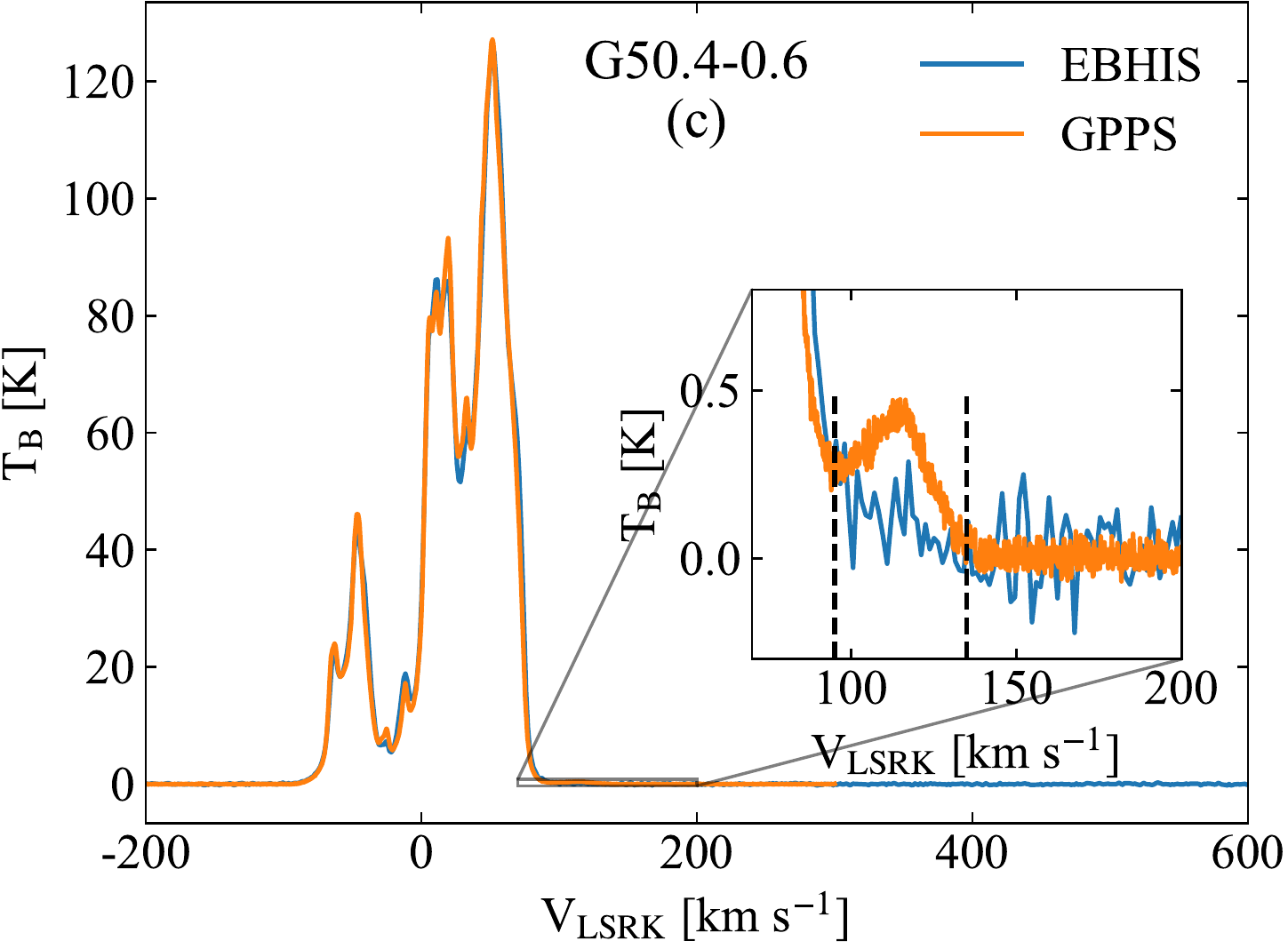} \\
	\hspace{0.1\textwidth} \indent 
	\includegraphics[width=0.31\textwidth]{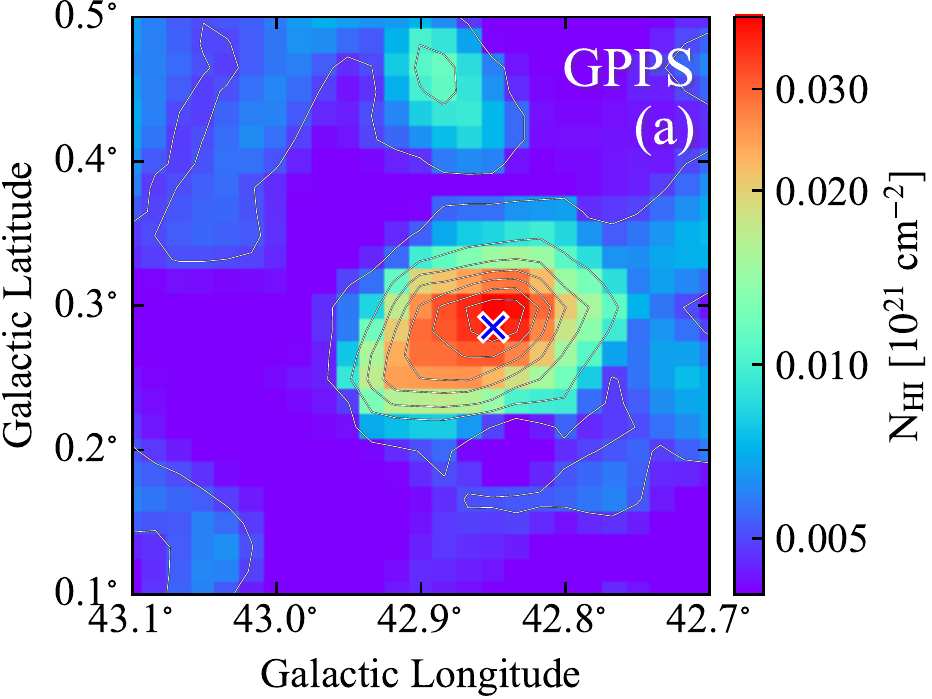}
	\includegraphics[width=0.31\textwidth]{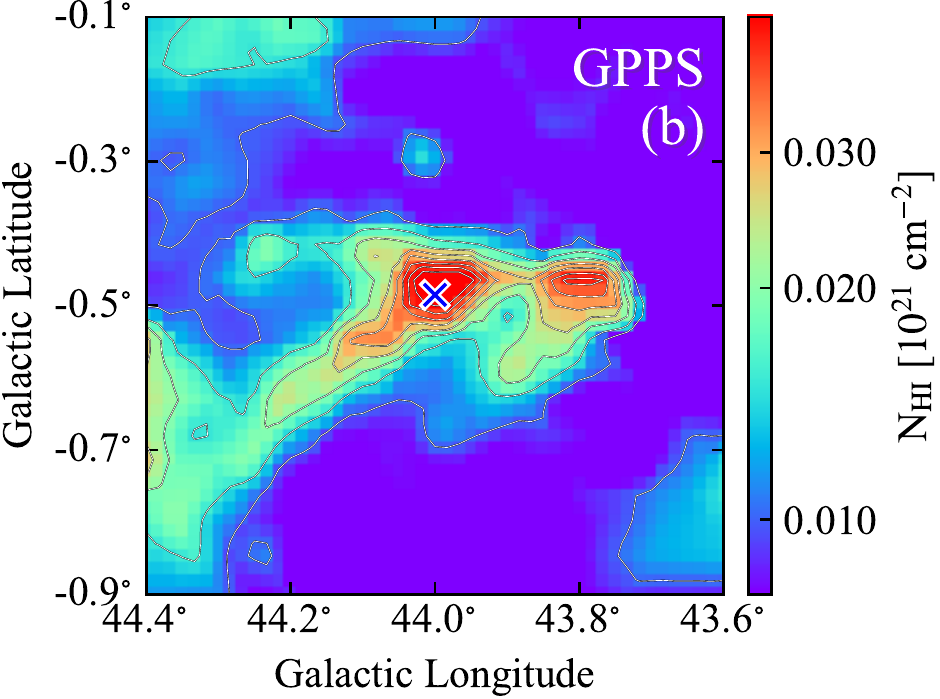}
	\includegraphics[width=0.31\textwidth]{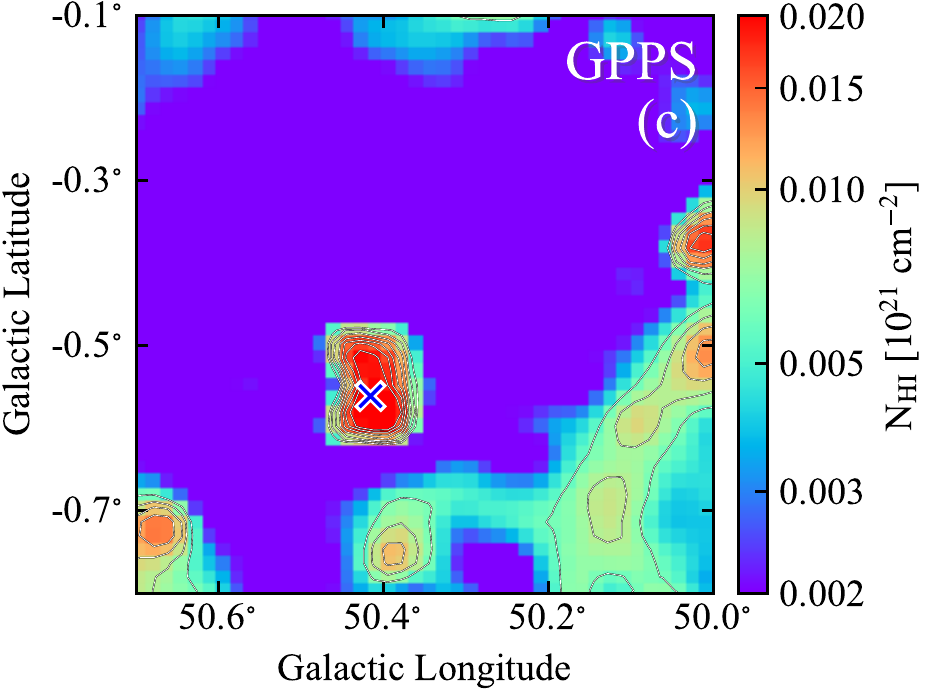}\\[-3.2mm]
	\hspace{0.1\textwidth}
	\includegraphics[width=0.31\textwidth]{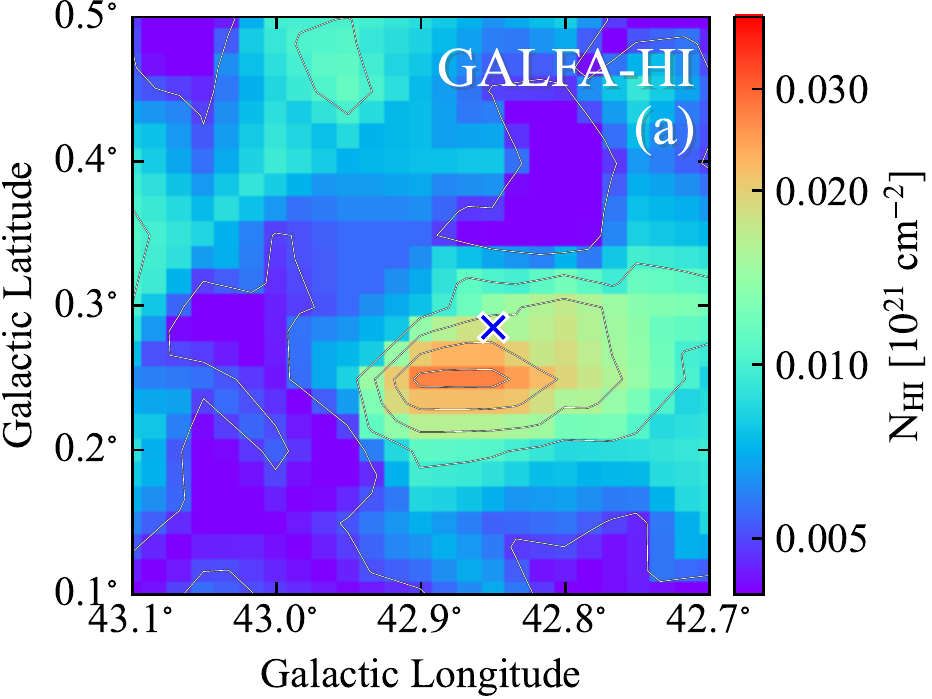}
	\includegraphics[width=0.31\textwidth]{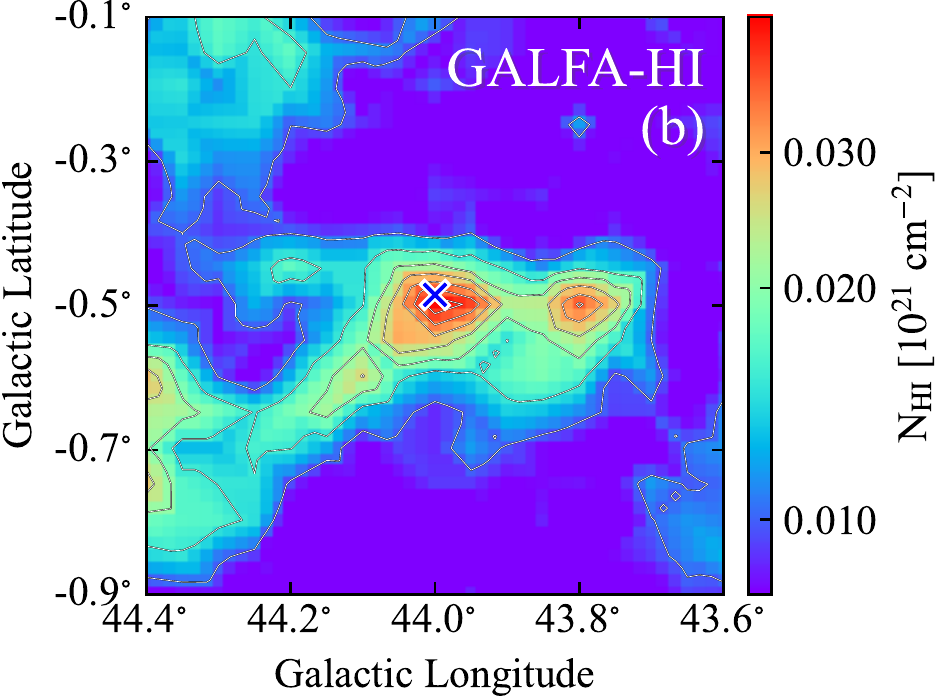}
	\includegraphics[width=0.31\textwidth]{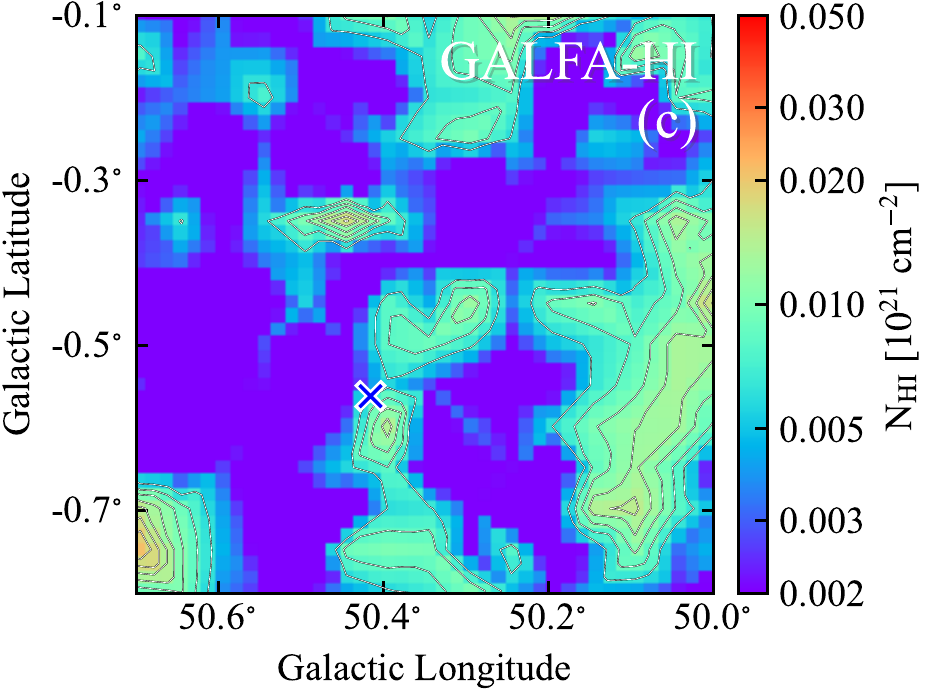} \\[-3.5mm]
	\hspace{0.1\textwidth}
	\includegraphics[width=0.31\textwidth]{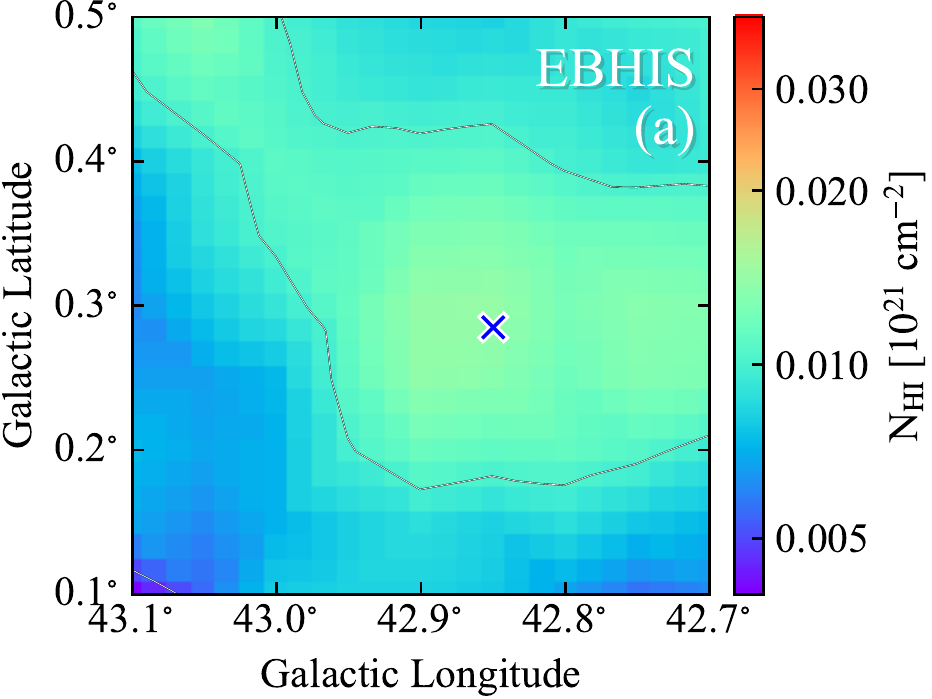}
	\includegraphics[width=0.31\textwidth]{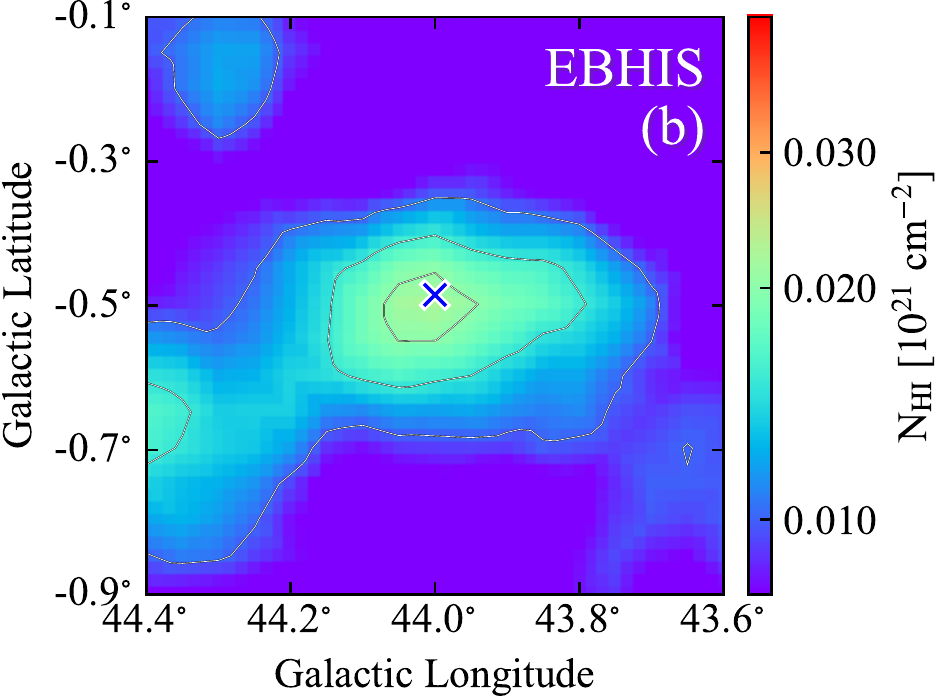}
	\includegraphics[width=0.31\textwidth]{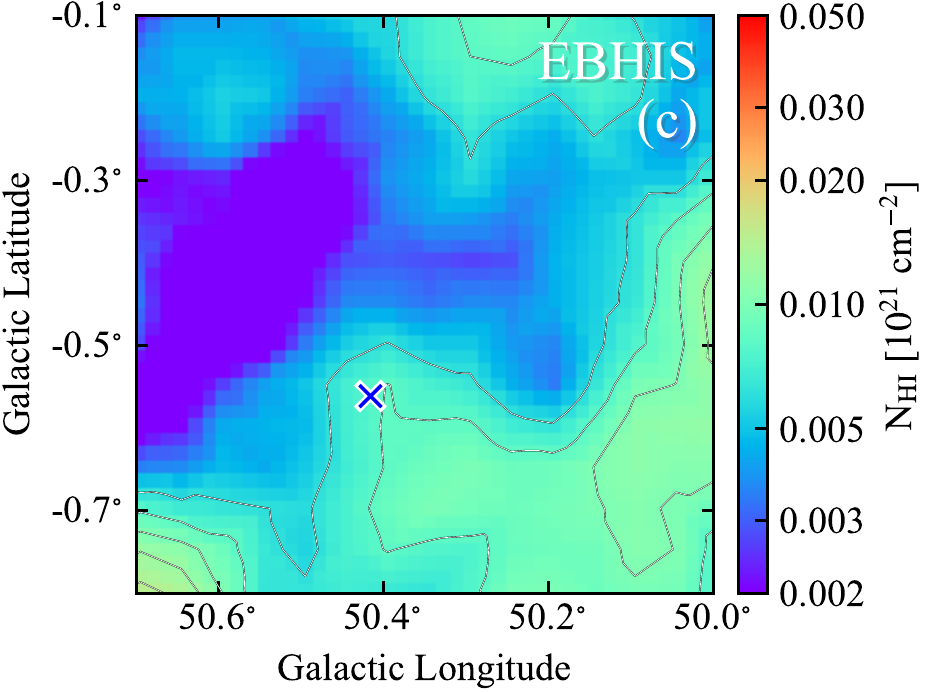} \\[-4mm]
	\hspace{0.1\textwidth}
	\includegraphics[width=0.31\textwidth]{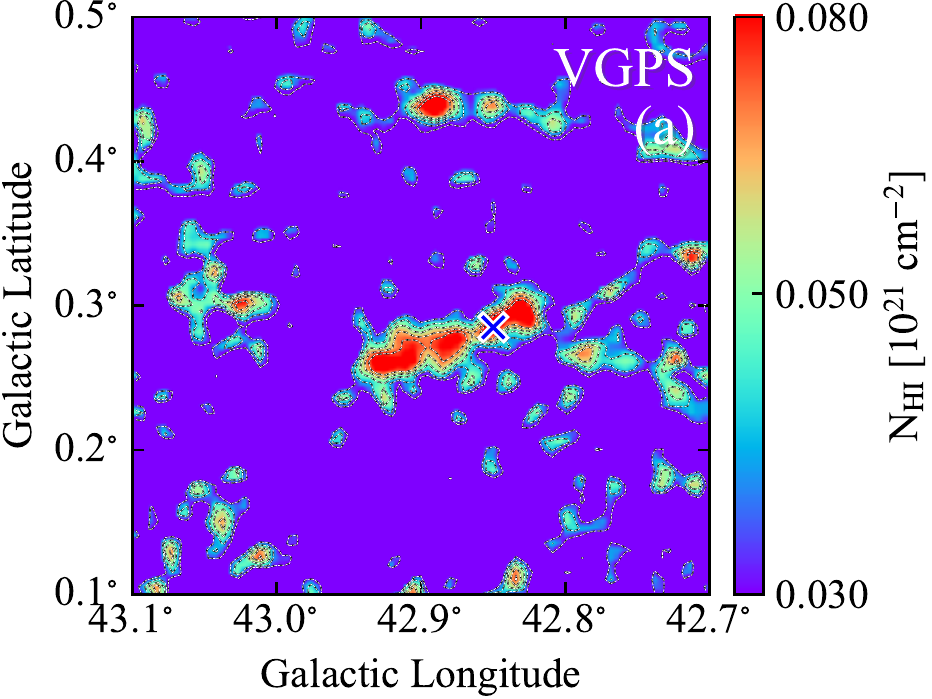}
	\includegraphics[width=0.31\textwidth]{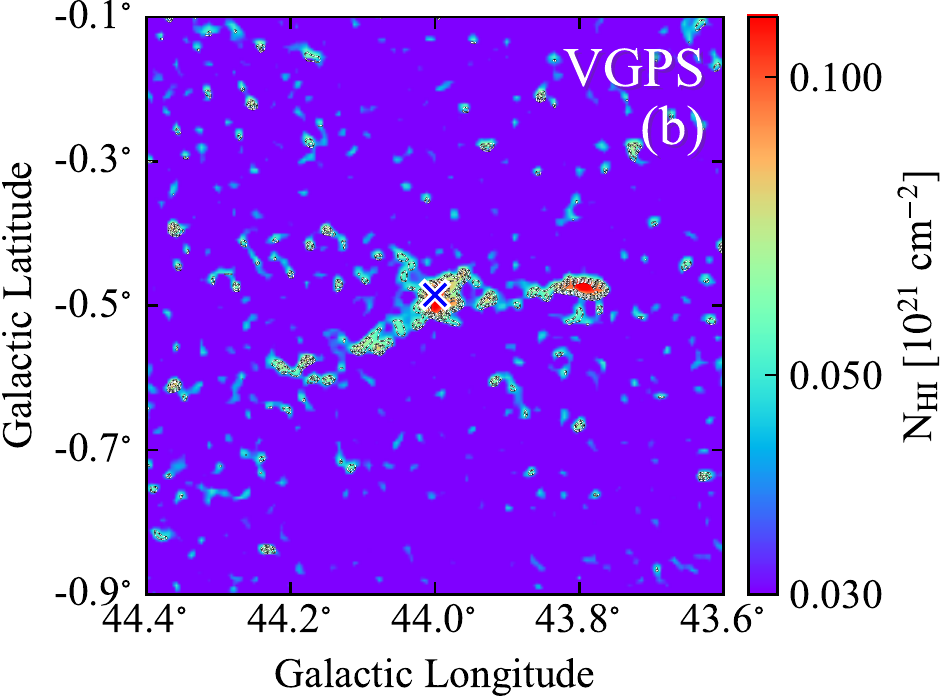}
	\caption{Three examples of weak exquisite \HI structures detected in the piggyback \HI data of the GPPS survey. The \HI spectra from the image peaks marked by ``$\times$'' in the panels of the second row are shown in the top panels by the orange lines together with the EBHIS spectra \citep[][]{Winkel2016} in blue lines, and the integrated column density images are shown in the second row of panels. The velocity range for the integrated images is indicated by the vertical dashed lines in the inserted enlarged plots of the top panels. For comparison, the integrated column density images from the GALFA-\HI \citep{Peek2018}, EBHIS \citep{Winkel2016} and the VGPS \citep{Stil2006} are shown in the panels of the third, fourth and the bottom rows, respectively. The exquisite \HI structure in the right panels is detected only in the GPPS survey, and it is too weak to be detected by the EBHIS (see the spectrum in the top panel) or the GALFA-H\textsc{i}. The VGPS data are polluted by RFI (image not shown). }
	\label{fig:weak_clouds}
\end{figure*}

\begin{figure*}[!t]
	\includegraphics[width=0.33\textwidth]{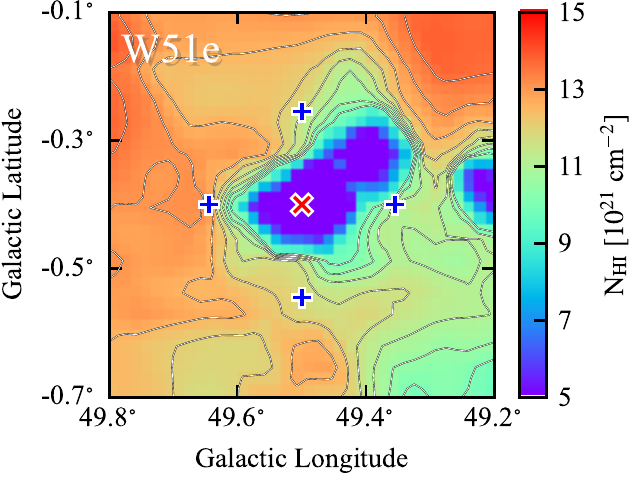}
	\includegraphics[width=0.33\textwidth]{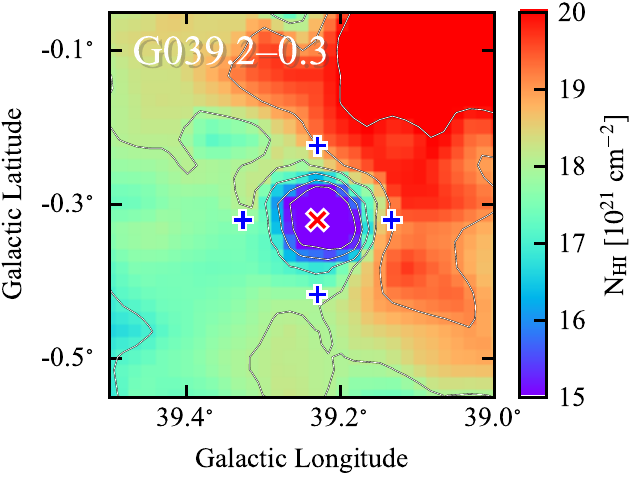}
	\includegraphics[width=0.33\textwidth]{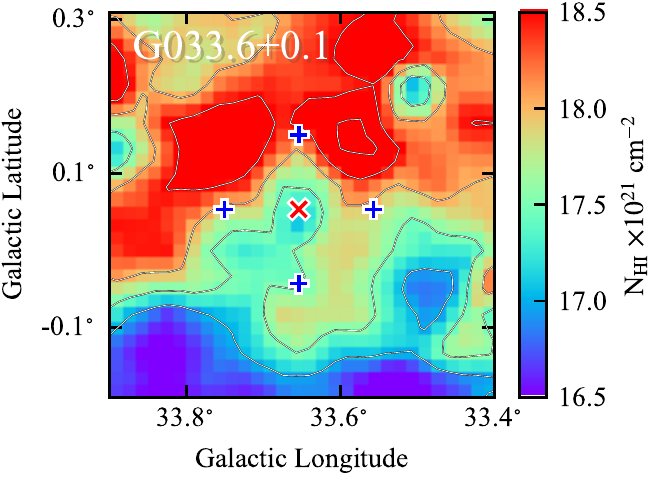}\\
	\includegraphics[width=0.33\textwidth]{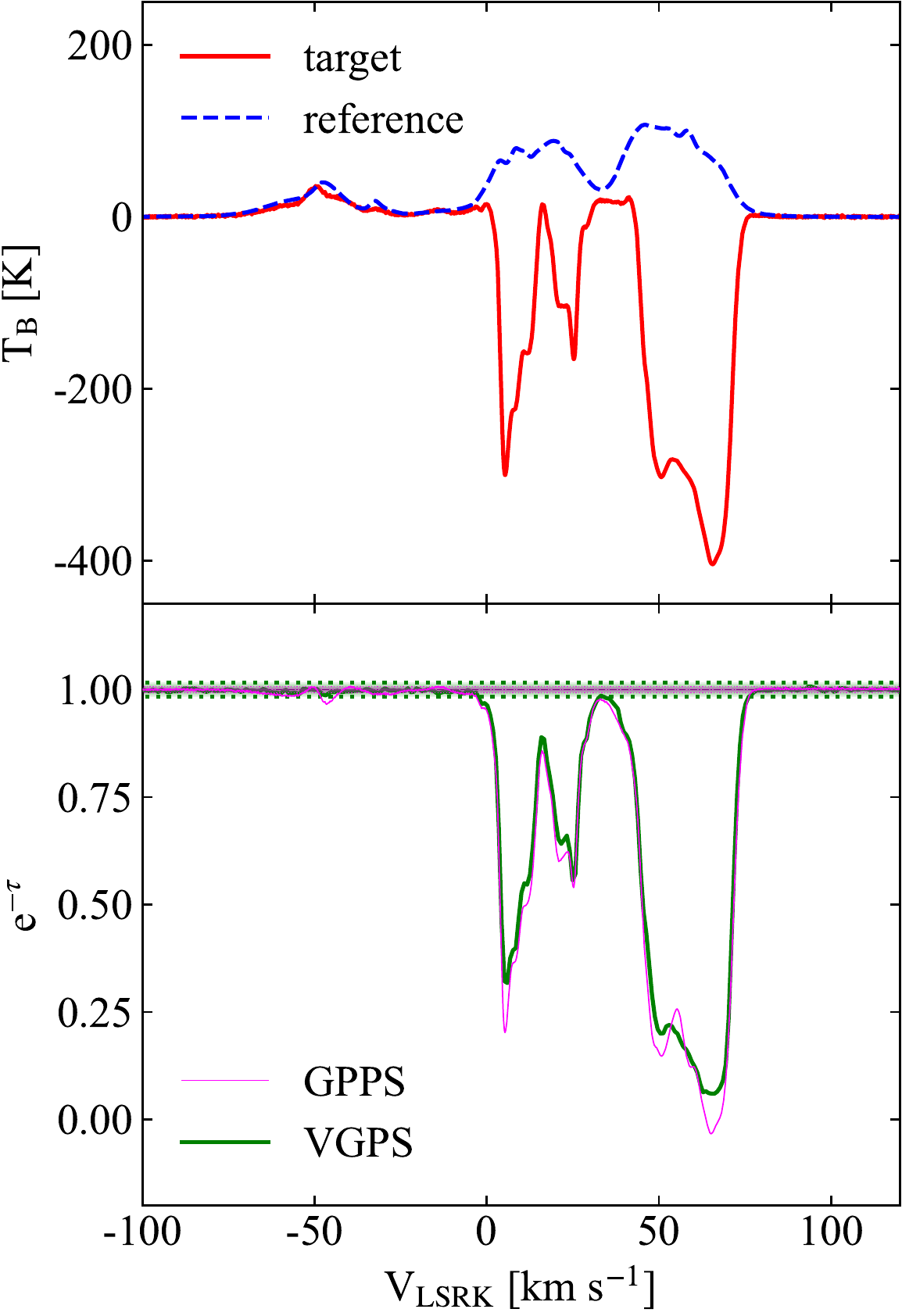}
	\includegraphics[width=0.33\textwidth]{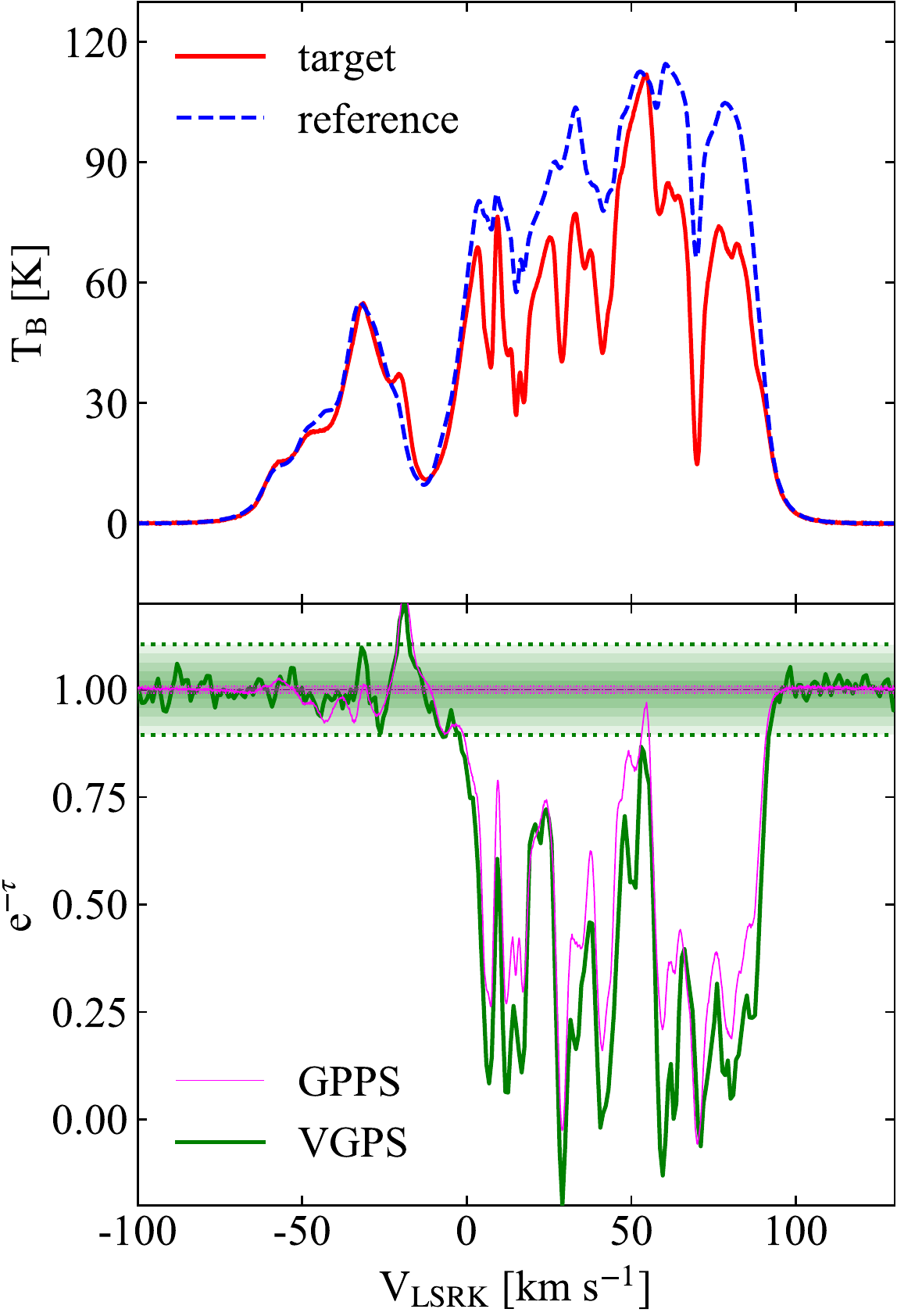} 
	\includegraphics[width=0.33\textwidth]{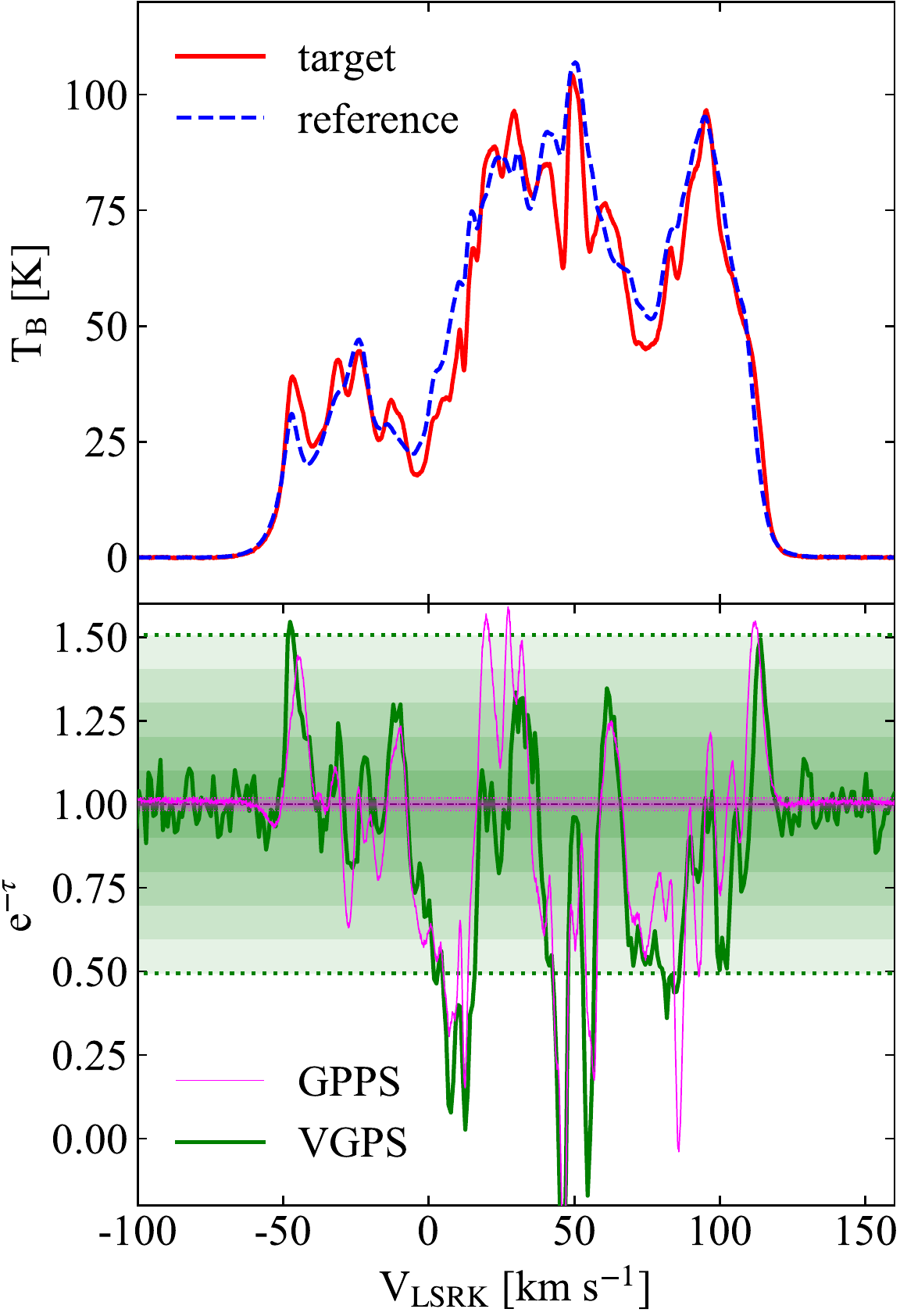} \\
	\caption{The \HI absorption features observed against W51e (left panels), SNR G039.2--0.3 (middle panels) and SNR G033.6+0.1 (right panels), respectively. The top panels show the integrated \HI column density map over the velocity range of $-150$~\kms\,to $150$~\kms\,centered at the targeted radio sources marked by the red ``$\times$''. The middle panels show the GPPS \HI spectra at the ``target'' (red solid line) and the averaged spectra from the four ``reference'' positions marked by ``+'' in the top map. The bottom panels show the \HI absorption spectra $e^{-\tau}$ extracted from GPPS \HI data (magenta solid line), together with those from the VGPS data \citep[green solid line,][]{Stil2006}, here $\tau$ is the optical depth. The shadow area and the magenta and green dotted lines indicate $\pm5\sigma$ area for the GPPS and VGPS spectrum, here the rms $\sigma$ is calculated from the line-free range of the spectra.}
	\label{fig:absorp}
\end{figure*}

\subsection{Capability to detect exquisite \HI structures}
		
The GPPS piggyback HI data have a high spatial resolution of $2.9'$ and a great sensitivity, and therefore weak and exquisite \HI structures can be well detected. Here we show three examples in Fig.~\ref{fig:weak_clouds} to demonstrate the quality of the GPPS \HI data. A huge amount of such structure details are left for readers to explore.
		
In the top panels of Fig.~\ref{fig:weak_clouds}, the very fine \HI spectra obtained from the GPPS survey data are presented for three peaks in the areas around ($l$, $b$) = ($42.9^\circ,\, 0.3^\circ$), ($l$, $b$) = ($44.0^\circ, \, -0.5^\circ$) and ($l$, $b$) = ($50.4^\circ,\,-0.6^\circ$),  together with the spectra from the EBHIS. Careful comparison of the line wings demonstrates that the faint clouds are detected by the GPPS survey. To show these clouds, the column density images are presented in the panels in the second row of Fig.~\ref{fig:weak_clouds}, which are made by the integration over the velocity ranges indicated by the vertical lines in the inserted panels for the enlarged spectra. Clearly the exquisite \HI structures are shown in unprecedented details, not only with elongated structures but also some compact cloud cores. 
		
To verify if these structures are real, we extract the \HI data from the EBHIS \citep{Winkel2016}, {GALFA-\HI \citep{Peek2018} } and VGPS \citep{Stil2006} for comparisons. The column density images are obtained by integrating in the same velocity ranges, and they are shown in the panels of the third, fourth and bottom rows of Fig.~\ref{fig:weak_clouds}. For all three regions, the EBHIS \HI maps \citep{Winkel2016} confirm these bright structures or cloud cores in a smoothed style due to its larger beam size. The GALFA-\HI maps have a spatial resolution of $4'$, it resolves the three structures better than the EBHIS data, and again confirm the FAST GPPS detection. In the high resolution \HI column density maps from the VGPS \citep{Stil2006}, the compact cloud cores are detected for the regions of ($l$, $b$) = ($42.9^\circ,\, 0.3^\circ$) and  ($l$, $b$) = ($44.0^\circ, \, -0.5^\circ$), and the map for the region of ($l$, $b$) = ($50.4^\circ,\,-0.6^\circ$) is severely affected by the RFI. The maps from the EBHIS, GALFA-\HI and VGPS therefore confirm that the detection of both extended structures and compact cloud cores by the FAST GPPS \HI data are reliable. 
The FAST GPPS \HI data have a better resolution and a better sensitivity 
than GALFA-H\textsc{i} (see Table~\ref{tab:comp}).
		
Using the kinematic model of the Milky Way developed by \cite{Reid2014}, the distances of these three exquisite \HI clouds can be estimated. The kinematic distances of \HI clouds may have a large uncertainty, and also suffer from the near and far side ambiguity inside the solar circle. The \HI cloud at ($l$, $b$) = ($44.0^\circ, \, -0.5^\circ$) in the middle column of Fig.~\ref{fig:weak_clouds} is located at an extreme negative velocity of $V_{\rm LSRK} \sim -81$ \kms, we estimate the kinematic distance of this \HI cloud as being approximately $21.1$ kpc. The clouds at ($l$, $b$) = ($42.9^\circ,\, 0.3^\circ$) and  ($l$, $b$) = ($50.4^\circ,\,-0.6^\circ$) have large positive velocities of $V_{\rm LSRK} \sim 110$ and 114 \kms, much larger than the possible velocities at the tangents of 73.4 and 53.8 \kms, respectively. The peculiar velocities of the cloud cores must be larger than 36.6 and 60.2 \kms, respectively.

In the region around  ($l$, $b$) = ($50.4^\circ,\, -0.6^\circ$), the GPPS \HI spectrum of the \HI image peak shows a clear bump in the velocity range of 100~\kms\,to 135~\kms\, with a peak brightness temperature of only 0.18 K, which is buried in the noise on the EBHIS spectrum. This GPPS integrated \HI image probably presents one of the weakest exquisite \HI structures in our Milky Way so far. 
		
\subsection{\HI absorption features}
\label{sec:absorp}
		
The most significant low column density spots in Fig.~\ref{fig:final} are the star formation complexes: W~49 and W~51; the supernova remnants: G034.7-00.4, G039.2-00.3 and G041.1-00.3, and the strong \HII region G034.3+00.1. 
{When observing the \HI against bright continuum sources, absorption is produced by neutral medium \citep{Dickey1990}. 
Continuum emission was subtracted in our data processing by the ArPLS \citep{Baek2015, Zeng2021} 
as the baseline in data.
Comparing with observations by using interferometers which can filter the unrelated surrounding diffuse emission, the results of single dish observations
contain the emission and absorption of the source and also the emission of surrounding gas. 
Nevertheless, the high sensitivity of the GPPS \HI data still provide a great chance for \HI absorption line studies. In Fig.~\ref{fig:absorp}, we show three examples for the \HI absorption lines, one star-formation region and two SNRs. The \HI absorption lines of other sources can be extracted by readers for their interests.
}

W~51 is one of the most active star-forming complexes in the Milky Way. It appears to be the lowest \HI column density region in Fig.~\ref{fig:final}. We extract the \HI absorption spectrum of W51e by selecting the center of the region as the target point (red cross in the top left panel of Fig.~\ref{fig:absorp}), the reference spectrum is the averaged spectrum from four points close to the target point but without strong continuum emission (blue crosses in the top left panel of Fig.~\ref{fig:absorp}). By comparing the `target' and `reference' spectra, we obtain the \HI absorption spectrum $e^{-\tau}$ as shown in the bottom left panel (magenta solid line), together with the \HI absorption spectrum extracted from the VGPS data (green solid line) using the same target and reference positions. The magenta and green dotted lines and shadows indicate the $\pm5\sigma$ area of two spectra, respectively. As one of the strongest continuum sources, almost the same absorption features are detected in the GPPS and VGPS \HI data in the positive velocity range. Because of the high sensitivity and velocity resolution, GPPS-\HI data reveal some sharper absorption features e.g. at the velocity of $V_{\rm LSRK} \sim -46.5$ \kms.
		
SNR G039.2--0.3 (3C~396) is a young SNR, appearing as a low density spot on the bright \HI emission region in Fig.~\ref{fig:final} and the top middle panel of Fig.~\ref{fig:absorp}. The \HI absorption spectra from the GPPS survey (this paper) and the VGPS data \citep{Stil2006} are shown in the middle column of Fig.~\ref{fig:absorp}. The rms of the GPPS spectrum is about 25 times smaller than that of VGPS data. In general, in the velocity range of $V_{\rm LSRK} = -10$ to 90 \kms\,the main absorption features of the two spectra are very similar, and the VGPS spectrum shows deeper absorption. The difference of absorption lines mainly comes from the different beam size of the two surveys, since a large telescope beam should have a beam smearing effect and hence weakens the absorption signal \citep{Dickey1990}. Another source of the difference is the filtering effect of interferometers. The GPPS \HI data contains the emission from large scale \HI structures but the VGPS does not. 
		
G033.6+0.1 is also an SNR. The absorption is obvious in the integrated \HI map in Fig.~\ref{fig:final} and the top right panel of Fig.~\ref{fig:absorp}, though it is among the  faint continuum sources. We obtain the \HI absorption spectra and show it in the right column of Fig.~\ref{fig:absorp}. The GPPS spectrum not only confirms the major VGPS absorption features, but also presents features at $V_{\rm LSRK} \sim 8, 12, 47, 55, 75, 83, 85$ and 100 \kms\,with a good confidence because of the high sensitivity.

\section{Summary}
\label{sec:summary}
		
The piggyback spectral data simultaneously recorded in the FAST GPPS survey \citep{Han2021} are valuable resources for the Galactic \HI studies. It has an angular resolution of  $2.9'$. With the five minutes integration time of each FAST pointing, the rms of the brightness temperature noise of the GPPS \HI data is approximately $40$~mK at a velocity resolution of $0.1$~\kms, which is the most sensitive survey of Galactic \HI emission by far.
		
We process the GPPS \HI data. A new routine is developed for automatically fitting and removing {the baseline and standing waves} efficiently. The \HI line intensity is scaled by using the standard on-off calibration {signals in each session. The} comparison of the GPPS integrated \HI intensity maps with the EBHIS survey gives the beam calibration factors for 19 beams and the cover correction factors for each cover, so that the long-term gain variations can be relatively calibrated. Because of the special observation mode of the GPPS survey, the released GPPS \HI data cube is not Nyquist-sampled in terms of the beam size. The correction for the stray radiation of different beams has not yet been carried out for the data set for this initial data release.
		
This first released GPPS \HI data are made for the survey area of $33^{\circ} \leq l \leq 55^{\circ}$ and $|b| \leq 2^{\circ}$, with a velocity resolution of $0.1$~\kms\,in the velocity range of $-300$~\kms\,$\leq V_{\rm LSRK} \leq 300$~\kms, which can be used to detect exquisite interstellar \HI structures and improve our understanding of the interstellar medium.

\section*{Data availability}
		
Original FAST GPPS survey data, including the piggyback spectral line data, will be released one year after observations, according to the FAST data release policy. All processed \HI data as presented in this paper is available on the project web-page: {\it \color{blue} http://zmtt.bao.ac.cn/MilkyWayFAST/}.
\newline
\vspace{5mm}
		
\footnotesize{\textbf{Acknowledgements.} 
This work has used the data from the Five-hundred-meter Aperture Spherical radio Telescope
(FAST), which is a Chinese national mega-science facility, operated by the National Astronomical Observatories of Chinese Academy of Sciences (NAOC). 
The GPPS project is one of five key projects carried out by using the FAST. 
The authors are supported by the National Natural Science Foundation (NNSF) of China No. 11988101, 11933011, 11833009,  the National Key R\&D Program of China (NO. 2017YFA0402701), the Key Research Program of the Chinese Academy of Sciences (Grant No. QYZDJ-SSW-SLH021) and the National SKA program of China No. 2022SKA0120103.  
%
TH is supported by the National Natural Science Foundation of China (No. 12003044). 
LGH thanks the support from the  Youth Innovation Promotion Association CAS. 
XYG acknowledges the support by the CAS-NWO cooperation programme (Grant No. GJHZ1865),
the National Natural Science Foundation of China (No. U1831103). 
CW was supported by NSFC No. 12133004. }

\bibliographystyle{raa}

\bibliography{bibfile}
		
\end{multicols}

\end{document}